%% file: edge3.tex
\documentclass[ASNA]{USG}

\usepackage{amsmath,amssymb,amsthm}
\usepackage{booktabs}
\usepackage{multirow}
\usepackage{graphicx}
\graphicspath{{figures/}{./figures/}{images/}{./images/}}
\makeatletter
\def\input@path{{tables/}{sections/}}
\makeatother
\usepackage{adjustbox}
\usepackage[table]{xcolor}

\providecommand{\qed}{\ensuremath{\square}}

\newcommand{\pibar}{\overline{\hat{\pi}}}
\newcommand{\DEFp}[1]{\textsf{EDGE-poly}\ensuremath{_{#1}}}
\newcommand{\DEFsym}{\textsf{EDGE-sym}}
\newcommand{\DEFstk}{\textsf{EDGE-stk}}

\newcommand{\EFname}{\textsf{HL}$_F$}

\articletype{}%
\received{}\revised{}\accepted{}
\journal{Preprint}
\copyyear{2026}\volume{00}\startpage{1}\articledoi{}

\makeatletter
\def\@@articletype{}%
\historydates{}%
\def\oddfoot@titlepage@info{\hfill{\pagenumfont\thepage\ of \@LastPg}}%
\def\evenfoot@titlepage@info{\oddfoot@titlepage@info}%
\makeatother
\usepackage{etoolbox}
\makeatletter
\patchcmd{\@maketitle}{\includegraphics[width=26.5mm]{Wiley_logo.eps}}{}{}{}
\patchcmd{\@maketitle}{\includegraphics[width=56mm]{allergy.eps}}{}{}{}
\makeatother

\begin{document}

\title{A grouped calibration test for logistic regression that tolerates a few corrupted records}

\author[1]{Ebrahim Khaled Ebrahim}[https://orcid.org/0009-0006-7839-8778]
\author[1]{Osama Abd El-Aziz Hussein}[https://orcid.org/0000-0002-6576-4912]
\author[1]{Ahmed El-Kotory}[https://orcid.org/0009-0000-1485-0768]
\authormark{EBRAHIM \textsc{et al.}}
\titlemark{A calibration test that tolerates corrupted records}

\address[1]{\orgdiv{Department of Applied Statistics}, \orgname{Alexandria University},
\orgaddress{\country{Egypt}}}

\corres{Ebrahim Khaled Ebrahim, Department of Applied Statistics, Alexandria University, Egypt.
\email{ebrahimkhaled@alexu.edu.eg}}


\abstract[Abstract]{The Hosmer--Lemeshow calibration test groups patients by predicted risk for
a valid reference and loses power that more groups cannot recover. Later tests weight each record's
residual alone, which makes them fragile: on an otherwise correct model, one record in a thousand with a
corrupted covariate raises the false-alarm rate of Stukel's score test from $5\%$ to $12\%$, and ten to
$63\%$.
Pooling on equal-size groups bounds a wrong record's influence, dividing its residual by its groupmates'
variance; without pooling, the same directions are as fragile as Stukel's test after a refit. EDGE, the test proposed
here, projects the grouped residuals onto a small basis of calibration shapes. Its degrees of freedom do
not depend on the number of groups, so the partition becomes a setting, fine-grained for power and
coarse when records may be wrong, and its reference needs no resampling. In a prespecified simulation
EDGE was comparable in power to Stukel's test on clean data and $0.064$ more powerful on average than the
ten-group Hosmer--Lemeshow test. With one and ten exaggerated covariates in a thousand its false-alarm rates were $0.056$ and $0.097$,
and with ten groups below $0.10$ up to about ten exaggerated or five reversed. In external validation, at ten groups, its average power was that of Cox's
recalibration test and the GiViTI belt, and its false-alarm rate stayed below $0.10$ with ten reversed
records in a thousand, where theirs did not. In a $70{,}000$-patient cohort its verdict did not change with the partition.}

\maketitle

\setcounter{secnumdepth}{3}

\input{sections/01_introduction}
\input{sections/02_test}
\input{sections/03_mechanism}
\input{sections/04_design}
\input{sections/05_power}
\input{sections/06_corrupted}
\input{sections/07_realdata}
\input{sections/08_discussion}
\input{sections/09_appendix}

\bmsubsection*{Author Contributions}
\textbf{Ebrahim Khaled Ebrahim:} conceptualisation; methodology; formal analysis; software;
investigation; data curation; visualisation; writing, original draft. \textbf{Osama Abd El-Aziz
Hussein:} supervision; validation; writing, review and editing. \textbf{Ahmed El-Kotory:} supervision;
project administration; validation; writing, review and editing. All authors read and approved the
final manuscript.

\bmsubsection*{Financial Disclosure}
This research received no specific grant from any funding agency in the public, commercial or
not-for-profit sectors.

\bmsubsection*{Conflict of Interest}
The authors declare no conflict of interest. The R package \texttt{ebrahim.gof}, which implements the
test proposed here and several of its comparators, is written and maintained by the first author and
is distributed free of charge under the GPL-3 licence.

\bmsubsection*{Ethics Statement}
This work required no ethical approval. It reports a simulation study together with an analysis of
the Diabetes 130-US Hospitals data \citep{clore2014uci}, a de-identified public data set released for
secondary research by the UCI Machine Learning Repository; no individual patient was contacted and no
identifiable data were used.

\bmsubsection*{Data Availability Statement}
No new patient data were generated. The clinical data analysed in Section~\ref{sec:realdata} are
publicly available from the UCI Machine Learning Repository \citep{clore2014uci} and are described by
\citet{strack2014impact}; the exclusion and preparation applied to them are in the archive below.

Everything else this paper reports is deposited in a public archive \citep{edge2_repro_2026}: the
study specifications with the SHA-256 hashes that fixed them before any result was computed, the
simulation code, the per-replicate $p$-values of every scenario, and the scripts that build every
table and figure.

The tests are implemented in the R package \texttt{ebrahim.gof} \citep{ebrahim2025gof}, available from
CRAN, where a single call to \texttt{run.all.gof()} runs the whole battery and returns one row a test,
and \texttt{run.all.external()} does the same for frozen predictions.

\bmsubsection*{Supporting Information}
Additional supporting information may be found online in the Supporting Information section at the
end of this article: S0, the comparisons specified before the study was run and their outcomes; S1,
the choice of weighting; S2, the three further corruptions; S3, the resampling tests under
corruption; S4, the robust fit in full; S5, partitions that pool only the extreme groups, and the
remedies that remove them; S6, the false alarms of every test under the exaggerated covariate;
S7, external validation against the usual tests; S8, EDGE on streaming data, with its proof; S9, power by departure family; S10, what
protects a test; S11, the proof of Theorem 2.

\bmsubsection*{Use of Artificial Intelligence Tools}
During the preparation of this manuscript, the authors used a large-language-model assistant
(Claude, Anthropic) to help draft and edit the wording and \LaTeX{} of the text. The authors
reviewed and edited all AI-assisted text and take full responsibility for the content of the
publication.

\bibliography{refs}

\end{document}

%% file: sections/01_introduction.tex
\section{Introduction}\label{sec:intro}

A clinical risk model is judged on two properties. Discrimination is whether it ranks patients
correctly, so that those who have the event were given higher risks than those who do not.
Calibration is whether the risk it reports is the risk the patient carries. A model can rank well and
still report the wrong number, and in medicine the number is what is used: a patient is treated,
screened or discharged when the predicted risk crosses a threshold. A model that predicts $8\%$ for
patients whose true risk is $5\%$ sends the wrong patients over that threshold, however well it
discriminates: discrimination asks only whether the model separates the patients at higher risk from
those at lower risk, not whether the risks it reports are right \citep{vancalster2016hierarchy}. Testing calibration is therefore routine in the
development and validation of clinical prediction models.

For logistic regression with continuous covariates the classical tests do not work. Almost every
patient has a covariate pattern of their own, so the Pearson and deviance statistics are sums of
cells with one observation each, and their chi-squared reference fails. Two repairs followed. One
standardises the Pearson statistic by its moments \citep{mccullagh1985,osius1992normal}. The other, by
\citet{hosmer1980goodness}, sorts the patients by predicted risk, cuts them into ten groups and
compares observed with expected events in each group. Grouping makes the cells large enough for a
reference distribution to exist, and the Hosmer--Lemeshow test is still the default check of
calibration.

Grouping has a cost: a data set of a million patients is reduced to ten numbers before the test sees
it, and the statistic spends one degree of freedom on each group, so its power is spread over every
direction in which a model can fail. Adding groups does not recover the lost information, because
each new group adds a degree of freedom faster than it adds signal, and the verdict can change from
rejection to acceptance with the number of groups alone \citep{paul2013power}. Most variants of the
test repair one of these problems and leave the others. \citet{paul2013power} and \citet{Lai2018}
standardise its power in large samples, \citet{nattino2020assessing} test whether the misfit exceeds a
size that does not grow with $n$, \citet{surjanovic2024generalized} derive a correct reference
distribution for a generalized form of the statistic, and BAGofT \citep{BAGofT2019} learns the
partition from the data instead of fixing it. \citet{hosmer2002goodness} came closest to a directed
grouped test: they weight each record's residual within the groups to gain power against a specified
alternative. None of them lets the partition be refined freely.

Other tests recover power by leaving the grouping behind. \citet{stukel1988generalized} embeds the
logistic link in a two-parameter family and tests the two extra parameters with a score test; in the
comparison of \citet{hosmer1997comparison} it had the best power against a misspecified link and was
among the tests they recommended. The GiViTI calibration belt \citep{nattino2014new,nattino2015new}
fits a polynomial to the calibration curve, and the Liu projection test \citep{liu2024comprehensive}
projects the individual residuals and refers them to a bootstrap. These tests use every record's
residual at its own weight, and that is why they are powerful. It also makes them sensitive to a few
wrong records. Clinical data contain such records: a covariate entered in the wrong units, a value
with a misplaced decimal point, a count typed as a larger but plausible count, which passes any range
check. With a model that is correct for every other patient, one record in a thousand with a
corrupted covariate raises the false-alarm rate of Stukel's score test from $5\%$ to $12\%$, and ten such
records raise it to $63\%$; in the hospital cohort of Section~\ref{sec:realdata}, about twelve miscoded counts among
$35{,}000$ patients raise it to $45\%$. A test that rejects for this reason is reporting a data-entry
error, and the analyst cannot tell it from a miscalibrated model; the two call for opposite actions,
cleaning the data or rebuilding the model.

Two tests keep a form of pooling without a fixed partition. The smoothed residual test of
\citet{le1991goodness} pools residuals over neighbourhoods in covariate space, and BAGofT pools them
over a learned partition. Le Cessie's test is protected against a few wrong records, and BAGofT showed
no rise in a small study (Section~\ref{sec:contamination}), but both pay in computation. Le Cessie's test builds an
$n\times n$ kernel, which on the $35{,}000$ patients of each half of the cohort of
Section~\ref{sec:realdata} would need about $10$ gigabytes, so it was not run there. BAGofT refits the
model more than ten thousand times and takes $392$ seconds for one data set of a thousand patients.

We keep the grouping and project the grouped residuals onto a small
basis of smooth calibration shapes, chosen to include the link departures that Stukel's test targets;
the weights act on groups, never on a single record. The statistic then spends a fixed number of
degrees of freedom however many groups there are, so the partition can be refined as the sample grows
without diluting the test, and its reference distribution needs no resampling or refitting: the test
takes $4$ milliseconds at $n=1000$ and half a second at $n=100{,}000$. We call the test EDGE (Efficient Directed Grouped Examination), the name
it has in the software. Its contributions are:

\begin{enumerate}
\item \textbf{A mechanism.} Pooling on equal-frequency groups bounds the influence that one corrupted
record's covariate can have on a goodness-of-fit statistic, because the record's residual is divided by
the variance of its clean groupmates (Proposition~\ref{prop:bounded}); the bound is exact in external
validation, where a frozen model is checked on new patients. A test loses this protection when it divides
a record's contribution by something that the record or the refit can make small: its own fitted
variance, or the information that the test's direction keeps after the fit. Stukel's test, the cubic
calibration test, an ungrouped test along EDGE's own directions and EDGE's score form all do, and break;
so do grouped tests whose extreme groups are small or that group in covariate space, and a robust fit or
a robust version of Stukel's test restores little
(Sections~\ref{sec:mechanism} and~\ref{sec:contamination}).
\item \textbf{A grouped test whose partition is a setting.} Because its degrees of freedom do not
depend on the number of groups (Theorem~\ref{thm:growingG}), the partition can be refined for power on
clean data and kept coarse for protection when records may be wrong
(Section~\ref{sec:partitionchoice}). In external validation it needs only outcomes and predicted
probabilities, so it applies to any model that reports them, logistic or not; there, at ten groups, its
average power is that of Cox's recalibration test and it keeps its false-alarm rate below $0.10$ where
that test does not. On clean data the test
is comparable in power to Stukel's score
test and the cubic calibration test; unlike them, it keeps its level when a few records are wrong, at
a modest cost in power.
\item \textbf{Evidence, including where the test loses.} A simulation study over six departure families
with every comparison and threshold fixed before it was run, twelve comparator tests including two
resampling tests and four other grouped tests, a study of external validation against the usual tests
there, and a cohort of $70{,}000$ patients in which
EDGE rejects at every partition while the Hosmer--Lemeshow verdict changes with the number of
groups.
\end{enumerate}

Section~\ref{sec:test} defines the test and its null distribution, and Section~\ref{sec:mechanism}
shows why pooling protects it. Section~\ref{sec:design} describes the simulation study,
Section~\ref{sec:power} reports power on clean data and Section~\ref{sec:contamination} under
corrupted records. Section~\ref{sec:realdata} applies the test to the hospital cohort, and
Section~\ref{sec:discuss} says when it is the right test and when another is better.

%% file: sections/02_test.tex
\section{The proposed test statistic}\label{sec:test}

\subsection{Grouped residuals and a basis of calibration shapes}

Sort the $n$ observations by fitted risk $\hat\pi_i$, ties in the order of the records, and cut them
into $G$ groups of equal frequency, as the Hosmer--Lemeshow test does. Write $O_g$ for the events observed in group $g$,
$E_g=\sum_{i\in g}\hat\pi_i$ for those expected, $V_g=\sum_{i\in g}\hat\pi_i(1-\hat\pi_i)$, $\pibar_g$
for the mean fitted risk of the group, and
\begin{equation}\label{eq:resid}
  r_g=\frac{O_g-E_g}{\sqrt{V_g}},\qquad r=(r_1,\dots,r_G)^\top .
\end{equation}
The Hosmer--Lemeshow statistic is $\lVert r\rVert^2$. It responds to a departure in any of the $G$
directions of residual space and spends a degree of freedom on each.

Suppose the fitted model is wrong through a misspecified link or an omitted smooth term, so that the
true event rate among patients of predicted risk $\pibar_g$ is $\pibar_g+\delta(\pibar_g)$ for a smooth
function $\delta$. Then $\mathbb{E}[r_g]\approx|g|\,\delta(\pibar_g)/\sqrt{V_g}$ is itself a smooth
function of $\pibar_g$, and a smooth function sampled at $G$ ordered points is described almost
entirely by a few shape components: a tilt, a bend and a second bend. The mean of $r$ therefore lies
close to a low-dimensional subspace, while the Hosmer--Lemeshow statistic weights all $G$ directions
equally.

Let $Z\in\mathbb{R}^{G\times d}$, with $d$ small, have columns that are smooth functions of the
calibration position. The test statistic is the squared length of the projection of $r$ onto the
span of $Z$,
\begin{equation}\label{eq:stat}
  S=r^\top Z(Z^\top Z)^{-1}Z^\top r=\lVert P_Z r\rVert^2 ,
  \qquad P_Z=Z(Z^\top Z)^{-1}Z^\top .
\end{equation}
The choice $Z=I_G$ returns the Hosmer--Lemeshow statistic, and the single column
$a_g=(1-2\pibar_g)/\sqrt{V_g}$ returns, up to scaling, the first-order correction of
\citet{osius1992normal} and \citet{farrington1996}. The groups and the residual vector are those of
the Hosmer--Lemeshow test; only the final combination differs. The weights act on groups, through the
basis, and never on the residual of an individual record, which is what Section~\ref{sec:mechanism}
turns on.

\subsection{The bases}\label{sec:bases}

Three bases are used, each a choice of alternative rather than a tuning parameter. With
$\bar\eta_g=\operatorname{logit}(\pibar_g)$:
\begin{description}
\item[\DEFp{3}, the default.] Orthogonal polynomials $P_1,P_2,P_3$ of $\pibar_g$, the tilt, bend and
double bend of the calibration curve; $d=3$.
\item[\DEFstk{}.] The grouped form of Stukel's constructed variables, $\bar\eta_g$,
$\bar\eta_g^2\mathbf 1\{\bar\eta_g\ge0\}$ and $-\bar\eta_g^2\mathbf 1\{\bar\eta_g<0\}$, which let the link
bend separately in each tail; $d=3$. It is a grouped version of Stukel's score test
\citep{stukel1988generalized}.
\item[\DEFsym{}.] The single column $\bar\eta_g|\bar\eta_g|$, the symmetric member of Stukel's
family, in which both tails bend together; $d=1$.
\end{description}
We write EDGE for the test in general and name the basis where it matters.

The columns of $Z$ may be used as they stand, the \emph{unit} form, or first scaled by $\sqrt{V_g}$,
the \emph{score} form, in which $Z^\top r$ is the score for adding the grouped shape to the model. The
score form moves weight towards the extreme risk groups. The unit form with the cubic basis is the
default and is used throughout unless stated. With the symmetric basis the score form is the better
choice when discrimination is high (an area under the curve near $0.9$ in our designs) and when a
covariate is strongly skewed, provided the data are clean: because it leans on the extreme groups, it
is the form most affected by corrupted records (Supporting Information S1).

\subsection{The null distribution}\label{sec:null}

The model is fitted to the same data, so $r$ is not a vector of independent standard normals. With
$X$ the model matrix, $W$ the fitted weights and $U$ the $G\times p$ matrix with rows
$U_g=\sum_{i\in g}\hat\pi_i(1-\hat\pi_i)x_i^\top/\sqrt{V_g}$,
\begin{equation}\label{eq:omega}
  r\ \dot\sim\ N(0,\Omega),\qquad \Omega=I_G-U(X^\top WX)^{-1}U^\top .
\end{equation}
Fitting the model absorbs part of each group's residual, so the diagonal of $\Omega$ is below one.
When a model is validated on new data with its coefficients frozen, nothing is estimated from those
data and $\Omega=I_G$. No score equation then absorbs the overall level, so a constant column is added
to the basis and the statistic has $d+1$ degrees of freedom; we call this the external mode. It uses
only the outcomes and the predicted probabilities, so it applies to the predictions of any model,
logistic or not, provided they were produced without the validation data.

\begin{theorem}[Null distribution]\label{thm:null}
Under the null hypothesis and conditions (A1)--(A4) of Appendix~\ref{app:proofs},
\begin{equation}\label{eq:wchi2}
  S\ \dot\sim\ \sum_{j=1}^{d}\lambda_j\,\chi^2_{1,j},\qquad
  \{\lambda_j\}=\text{eigenvalues of }(Z^\top Z)^{-1}Z^\top\Omega Z ,
\end{equation}
with $0\le\lambda_j\le1$ for every $j$.
\end{theorem}

The $p$-value is computed from the Satterthwaite scaled chi-squared approximation to this weighted
sum \citep{satterthwaite1946}, which is exact when $d=1$; Imhof's numerical inversion \citep{imhof1961}
is available when more accuracy is wanted. Every simulation in this paper uses the Satterthwaite
approximation. Neither computation needs a second model fit or resampling.

\subsection{The partition as a setting}\label{sec:partitionchoice}

\begin{theorem}[Growing partition]\label{thm:growingG}
Let the basis columns be the values at the group positions of fixed, continuously differentiable,
linearly independent functions on $[0,1]$, and let $G=G_n\le n$ with $G_n/\sqrt{\log n}\to\infty$.
Under the null hypothesis and conditions (A2), (A3) and (A4$'$) of Appendix~\ref{app:proofs}, $S$
converges to a weighted sum of $d$ chi-squared variables on one degree of freedom whose weights, given
in \eqref{eq:growingG}, do not depend on $G_n$. No lower bound on the number of observations per
group is required.
\end{theorem}

The rate condition is mild: the default partition below grows linearly in $n$ and satisfies it.

The Hosmer--Lemeshow statistic spends one degree of freedom per group, so a finer partition dilutes
it. The test statistic spends $d$ degrees of freedom whatever $G$ is, so the number of groups
becomes a setting that the analyst chooses, and we use two. The \emph{default partition} refines with
the sample, $G=\max(10,\lceil n/25\rceil)$, about twenty-five patients a group (the simulation study
rounded $n/25$ to the nearest integer, which moves the partition by at most one group); on clean data it gives
EDGE more power than ten groups (Section~\ref{sec:verdicts}). When records may be wrong,
the \emph{ten-group partition} keeps many clean records in the extreme groups, where corrupted records
collect, and protects the cubic basis much further (Section~\ref{sec:edgefr}); the symmetric basis is
better protected at the default partition against reversed records, though not against many
exaggerated ones (Section~\ref{sec:severity}). Pooling only the two extreme groups protects no better
than ten groups (Supporting Information S5).

In the R package \texttt{ebrahim.gof} \citep{ebrahim2025gof} the test is
\texttt{edge.gof(fit)} for a fitted \texttt{glm}. It reports two rows: the verdict at the default
partition and, as a robustness check, the same test at ten groups; its degrees of freedom are the
effective $\nu$ of the Satterthwaite reference. \texttt{G = 10} alone gives the ten-group partition, \texttt{basis = "sym"} or \texttt{"stukel"} for the other bases and
\texttt{weights = "score"} for the score form. For frozen predictions,
\texttt{edge.gof(y, predicted\_probs = p, external = TRUE, G = 10)} computes the external-mode
statistic at ten groups, with $\Omega=I_G$ and the constant column added, and reproduces every statistic
of Section~\ref{sec:realdata} at its own $G$; \texttt{run.all.external(y, p)} runs it at ten groups and at
the default partition beside the other external-validation tests of Supporting Information S7
(version 2.9.0 onwards).

%% file: sections/03_mechanism.tex
\section{Why pooling bounds a corrupted record}\label{sec:mechanism}

Consider a record whose covariate was entered wrongly, for example multiplied by four. Its linear
predictor becomes extreme, so its predicted risk goes towards zero or one and it sorts into an extreme
risk group. Its outcome, however, was generated by the covariate the patient actually had. If that
outcome contradicts the recorded prediction, the record is a bad leverage point in the sense of
\citet{rousseeuwleroy1987}; if it agrees, it is a good one. This section shows what one such record can
do to a goodness-of-fit statistic, and why pooling limits it.

\subsection{The record in the fit and in the test}

The record's contribution to the likelihood score is $x_i(y_i-\hat\pi_i)$ and its contribution to the
information is $\hat\pi_i(1-\hat\pi_i)x_ix_i^\top$. At an extreme prediction the second is close to
zero. The first is close to zero when the outcome agrees with the prediction and of the order of
$\lVert x_i\rVert$ when it contradicts it, so a contradicting record pulls the fit, and with one such
record at an arbitrarily extreme covariate the logistic maximum-likelihood estimator breaks down
\citep{croux2002breakdown}. In our designs a covariate multiplied by four changes the fitted slope by
less than $5\%$ with ten corrupted records in $1000$; a sign error, which reverses the prediction and so
makes most corrupted records bad leverage points, changes it by $14\%$.

A corrupted record sits alone at an extreme of the predicted risk. A test is fragile when it divides
that record's contribution by something the record, or the refit it causes, can make small. One such
divisor is the record's own fitted variance $\hat\pi_i(1-\hat\pi_i)$, which vanishes at an extreme
prediction. Another is the information that the test's direction keeps after the fit: a smooth
direction close to the fitted linear predictor keeps little, so a score test that divides by it
magnifies whatever the corrupted record and the slightly moved fit leave along that direction. We call
tests that standardise each record separately \emph{record-level}; Stukel's score test, the cubic
calibration test and the GiViTI belt are of this kind. A test is \emph{pooled} when it combines each record's residual with those
of other records, over groups of risk, over a learned partition or over neighbourhoods in covariate
space, and standardises the sum by the variance of the whole pool.

In the grouped residual \eqref{eq:resid} the corrupted record acts in two ways. Let $\theta$ be the
probability that its outcome contradicts its recorded prediction. In our designs, for a covariate
multiplied by four, $\theta$ is about $0.3$ over all corrupted records and $0.1$ among those that reach
the top group; for a sign error it is about $0.7$ and $0.8$. The record shifts the
numerator: at the top of the risk scale, where $\hat\pi_i\approx1$, its expected contribution to
$O_g-E_g$ is about $-\theta$, and at the bottom about $+\theta$. It also leaves $V_g$ understating the
variation it brings, because $V_g$ is computed at $\hat\pi_i(1-\hat\pi_i)\approx0$ while the record
varies with its true risk. With $m_c$ corrupted records among the $m$ records of a group, $\bar v$ the
mean fitted variance of a clean member and $\bar v_c$ the mean true variance of a corrupted one, a
heuristic first-order approximation gives
\begin{equation}\label{eq:twomoments}
  \mathbb{E}[r_g]\;\approx\;\mp\frac{m_c\,\theta}{\sqrt{(m-m_c)\,\bar v}},
  \qquad
  \operatorname{Var}(r_g)\;\approx\;1+\frac{m_c\,\bar v_c}{(m-m_c)\,\bar v},
\end{equation}
with the minus sign in the top group and the plus sign in the bottom one. In both expressions the
corrupted records are in the numerator and the clean groupmates form the whole denominator: the
contribution of one corrupted record among twenty-four clean ones is divided by the square root of
their summed variance.

\subsection{Bounded influence at fixed coefficients}

\begin{proposition}[Bounded influence in the covariates]\label{prop:bounded}
Hold the coefficients at $\hat\beta$. Replace one record's covariate by an arbitrary value $x^{*}$,
leaving its outcome and every other record unchanged, and write $\eta^{*}=x^{*\top}\hat\beta$. For an
ungrouped directed score test with direction $z(\eta)=\eta|\eta|$ the statistic grows without bound as
$\eta^{*}$ moves in the direction in which the record's outcome contradicts its prediction: the record's contribution to the score is of order $\eta^{*2}$, while its contribution to
the information, $z(\eta^{*})^{2}\hat\pi^{*}(1-\hat\pi^{*})$, tends to zero. For the grouped statistic,
if the record stays in its group and $V_g>1/4$,
\[
  |\Delta r_g|\;\le\;\Bigl(V_g-\tfrac14\Bigr)^{-1/2}
  \Bigl\{1+\frac{|r_g|\sqrt{V_g}}{8\,(V_g-\frac14)}\Bigr\},
\]
and if it changes group the same bound holds with $2$ in place of $1$ for every group it touches.
None of these bounds involves $\eta^{*}$.
\end{proposition}

The proof is in Appendix~\ref{app:proofs}. The proposition is exact in external validation, where
the coefficients are frozen and nothing is refitted. When the model is refitted to the contaminated
data it bounds the record's direct influence on the statistic, not its influence through
$\hat\beta$, which is that of the maximum-likelihood estimator; the measured slope changes above say
how large that is in our designs. This is a finite-sample bounded-influence property in the spirit of
\citet{hampel1986robust}, not a statement about the asymptotic influence function.

The bound is small only when $V_g$ is large. At the default partition an extreme group holds about
twenty-five records whose mean risk is near zero or one, so $V_g$ can be close to $1/4$ and the bound is of
order one; that is where the default partition loses its protection (Section~\ref{sec:severity}), and
ten groups, with $V_g$ in the tens, restore it. The bound holds for any statistic built from grouped
residuals, the Hosmer--Lemeshow statistic included; EDGE inherits it and spends the grouped residuals on
a few directions with unit weights, so that nothing in its denominator depends on the fit's remaining
information. Bounded directions alone do not give the bound its force: Proposition~\ref{prop:bounded}
concerns a record's direct contribution at fixed coefficients, and a score test for the same cubic
directions in the predicted risk passes that test too, yet once the model is refitted it is as fragile
as Stukel's test, and so is EDGE's own score form (Section~\ref{sec:whatprotects}).

Any form of pooling supplies the divisor: le Cessie's test pools over neighbourhoods in covariate
space and is protected (Section~\ref{sec:contamination}). For grouping on the risk, the protection is
set by how many clean records share the groups where corrupted records land, and so by the
partition rather than the contamination rate. The Hosmer--Lemeshow statistic with equal-width rather
than equal-frequency risk groups shows this directly: its extreme groups hold few records, and one
corrupted record in a thousand raises its false-alarm rate to $0.151$ (Table~\ref{tab:severity}). The
protection ends when corrupted records fill a group, because as $m_c$ approaches $m$ the denominator
empties. The heuristic describes the cubic basis well; how the symmetric basis responds to the
partition under a sign error, which Section~\ref{sec:severity} reports, it does not explain.

The bound concerns records whose \emph{prediction} is wrong. A record whose \emph{outcome} is wrong,
for example an event that was never recorded, changes $O_g$ in the direction a miscalibration would,
and a calibration test should detect it.

%% file: sections/04_design.tex
\section{Simulation study}\label{sec:design}

Every scenario, comparison, threshold and decision rule below was written down and fixed, with a
cryptographic hash, before any of its results were computed. The specifications, their hashes and
every per-replicate $p$-value are in the reproducibility archive \citep{edge2_repro_2026}; the
comparisons that were specified in advance, and their outcomes, are listed in Supporting Information
S0. Section~\ref{sec:reading} lists where the analysis departs from them.

\subsection{Designs and departures}\label{sec:families}

The working model is a logistic regression for a binary outcome with one to three covariates. Data are
generated from a truth that departs from it in one of six families:
\begin{enumerate}
\item \textbf{Symmetric tails}: the cauchit and $t_4$ latent links, and the symmetric members of
      Stukel's family, heavier or lighter than the logistic at both ends.
\item \textbf{Asymmetric links}: the complementary log--log and log--log links and the asymmetric
      members of Stukel's family, which bend the curve once.
\item \textbf{The Stukel symmetric family} at its published shape parameters.
\item \textbf{Omitted terms}: a quadratic term, an interaction, or a covariate left out of the model.
\item \textbf{Rough misfit}: an oscillating calibration curve, which no smooth basis can follow.
\item \textbf{Off-index departures}, which do not act through the linear predictor.
\end{enumerate}
Sample sizes run from $100$ to $50{,}000$. Each family is crossed with three regimes: a base design
(area under the curve about $0.77$), a high-discrimination design (about $0.89$), and a design with
$12\%$ events, the sparse regime in which partition tests are known to struggle. There are $131$ null
and $158$ alternative scenarios, with between $500$ replicates in the largest samples and $10{,}000$ in the
smallest.

\subsection{Comparators}\label{sec:comparators}

\begin{itemize}
\item The Hosmer--Lemeshow statistic and its Osius--Rojek/Farrington correction \EFname{}
      \citep{hosmer1980goodness,osius1992normal,farrington1996}, at $G=10$ and at the default
      partition; the Hosmer--Lemeshow statistic with equal-width risk groups
      \citep{hosmer1980goodness}; the Pigeon--Heyse test \citep{pigeon1999b}; and the tests of
      \citet{Tsiatis1980} and \citet{xie2008increasing}, which group the records into ten $K$-means
      clusters in covariate space, Tsiatis's through a score test on the cluster indicators. These four
      run at ten groups.
\item Stukel's score test in its joint and one-parameter forms, and its likelihood-ratio refit
      \citep{stukel1988generalized}. The form distributed in current software \citep{LogisticDx}
      refers the sum of two correlated score components to a chi-squared distribution on two degrees of
      freedom; that reference is not correct, with a median size of $0.066$ over the $154$ null settings in which it was run, those of the
      supplementary studies included
      (quartiles $0.049$ to $0.070$) against $0.049$ for the joint score, so the joint
      score form is used throughout.
\item The GiViTI calibration belt \citep{nattino2014new,nattino2015new}.
\item The cubic calibration likelihood-ratio test, which adds the square and cube of the fitted
      linear predictor to a logistic regression of the outcome on it and compares the fits on two
      degrees of freedom: the polynomial calibration model of the GiViTI belt with its degree fixed at
      three, which extends Pregibon's goodness-of-link test
      \citep{pregibon1980goodness,nattino2014new,nattino2015new}.
\item The kernel-smoothed residual test of \citet{le1991goodness,le1995goodness}, at its default kernel
      and bandwidth. We use the implementation in \texttt{ebrahim.gof}, because the implementation it
      was derived from omits a transpose that a weighted fit requires and is mildly liberal as a result
      \citep{ebrahim2026benchmark}.
\item The Liu projection test \citep{liu2024comprehensive}, a logistic form of the statistic of
      \citet{escanciano2006consistent}, with $B=250$ model-based bootstrap draws, from our implementation, which is in the archive and in
      \texttt{ebrahim.gof} from version 2.8.0.
\item BAGofT \citep{BAGofT2019} from its published package at its defaults: one hundred splits, one
      hundred simulations and the random-forest partition. One line of the package fails on models with a
      single covariate and was repaired; with two or more covariates the repaired function returns
      identical results.
\end{itemize}
The weighted grouped tests of \citet{hosmer2002goodness}, which multiply each record's residual by a
weight chosen for power against a specified alternative, are the closest earlier directed grouped
tests. Their weights are the residuals of a weighted regression of the alternative's direction on the
covariates, so they can grow with a record's covariates and Proposition~\ref{prop:bounded} does not
cover them; they are compared, with the paper's general-purpose weight, in the study of
Section~\ref{sec:whatprotects}. Every test in the list except the cubic calibration test is available in \texttt{ebrahim.gof} \citep{ebrahim2025gof}, where one call to
\texttt{run.all.gof()} fits the model once and returns all of them; \citet{ebrahim2026benchmark}
benchmark that battery under sparse data. The three tests outside the
partition family are slow, so they were run on subsets of the scenarios: le Cessie's test on $82$, the
projection test on $45$ and BAGofT on $32$. Every test sees the same data sets.

\subsection{Corrupted records}\label{sec:contamination-design}

In each replicate $k$ of the $n$ records are corrupted, with $k/n$ from $0.1\%$ to $1\%$ and $n$ equal
to $1000$ or $5000$:
\begin{description}
\item[C1, an exaggerated covariate.] The covariate of $k$ random records is multiplied by four, and
      their outcome is still drawn from the truth at the original value. The \emph{prediction} is wrong
      and the outcome is not.
\item[C1$'$, a sign error.] As C1, with the covariate multiplied by $-4$. The prediction is reversed
      rather than exaggerated.
\item[C2, missed events.] The $k$ records of highest true risk have their outcome set to zero. The
      prediction is right and the \emph{outcome} is wrong.
\item[C3, a unit error.] The covariate of $k$ records is divided by ten, which makes their
      predictions milder rather than extreme.
\item[C4, duplicated records.] The $k$ highest-risk records are copied over $k$ random rows, which
      checks that the grouped reference survives ties.
\end{description}
Formally, the null hypothesis is that the model is calibrated for the patients as they are,
$\Pr(y=1\mid x)=\pi(x;\beta)$ at the true covariates, while $k$ of the $n$ records are observed with
covariates $\tilde x\neq x$ and so receive the prediction $\pi(\tilde x;\hat\beta)$. Under a logistic truth
the model is correct for every uncorrupted record, so a rejection is a false alarm: the test is
reporting the corrupted records, not a miscalibration of the model for the population. It is not a type-I error in the strict sense, because a contaminated sample is a mixture; what
is measured is each test's stability of level over a small gross-error neighbourhood of the model
\citep{huber1964robust,hampel1986robust}. It is not the classical measurement-error problem, in which
every covariate is observed with error and the fit itself is biased \citep{carroll2006measurement};
here a few records carry a gross error and the fit moves by the amounts given in
Section~\ref{sec:mechanism}. Under a probit and a complementary log--log truth the same corruptions are
applied to a real misfit, to measure the cost in power. A value corrupted this way lies outside the
range of the clean covariate, and a range check on that variable would find it; the cohort of
Section~\ref{sec:realdata} shows an error that no range check finds. Wrong outcomes (C2) are treated differently: the recorded data are then miscalibrated, and a
calibration test should detect them. A further study compares EDGE with robust quasi-deviance versions
of Stukel's test \citep{cantoni2001robust}, and two others vary the partition with the data held fixed
and compare partitions that pool the extreme groups at contamination rates up to $4\%$. A last study, added after the main results were read and hashed before it ran, runs on the same data
sets EDGE, an ungrouped score test for the same
cubic directions in the predicted risk, EDGE with one record a group, Spiegelhalter's $z$
\citep{spiegelhalter1986probabilistic}, the weighted grouped tests of \citet{hosmer2002goodness}, and
each test after the records flagged by Pregibon's diagnostics \citep{pregibon1981logistic} are removed;
it adds misfit confined to the top $2\%$ of risk, to measure what protection costs, and a design with
twelve covariates at $n=2000$.

\subsection{Reading the results}\label{sec:reading}

Size is checked before power is reported. Power is size-adjusted: each test is read at the critical
value from its own matched null scenario, so every test is compared at its own realised level; where
a table reports raw rejection rates, it says so. Comparisons are paired: two tests are compared on the
same replicates by an exact McNemar test on the discordant pairs, with Holm's correction within each
family of comparisons. A test \emph{leads or ties} a rival in a scenario when its size-adjusted power
is no more than $0.01$ below the rival's; the specification allowed $0.014$, and the stricter margin is
used everywhere. A replicate on which a test returns no $p$-value counts as a non-rejection. The two
resampling tests return $p$-values on grids of $1/100$ and $1/250$ and reject at $p<\alpha$; every other
test rejects at $p\le\alpha$. Under corruption we say a test \emph{holds} while its false-alarm rate
stays below $0.10$, twice the nominal level; that is the bar the specification set. It is looser than
the band of three Monte Carlo standard errors used for size, and the tables give every rate so that
the reader can apply either.

Three readings depart from the specifications. The robust-fit study (Section~\ref{sec:robustest})
reads the contaminated columns of the tests whose level survived the robust fit, although its rule was
to read none once any test failed that check. The corruption experiment on the cohort was repeated on
a second covariate, chosen by a rule fixed before it was run, after the first showed no effect. And
the cohort analysis of Section~\ref{sec:realdata} keeps one admission per patient and splits by
patient, where the specification split admissions; the change was made because patients recur across
admissions.

%% file: sections/05_power.tex
\section{Power on clean data}\label{sec:power}

\subsection{Level and choice of basis}\label{sec:size}\label{sec:basis}

Table~\ref{tab:size} gives the size of each basis and weighting over the $131$ null scenarios. The
median is $0.048$ to $0.050$ in every form and the range is $0.013$ to $0.064$. In the unit form the
cubic basis falls outside a band of three Monte Carlo standard errors in one scenario of $131$, at the
$5\%$ level, the symmetric basis in one, at the $1\%$ level, and the Stukel basis in three, all at
$1\%$. The score form is less well behaved, failing four scenarios with the cubic basis and ten with
the Stukel basis. The rule fixed in advance therefore selected the unit form for all three bases.

\input{tab_size}

Table~\ref{tab:families} gives the mean size-adjusted power of the three bases in each departure
family. No basis leads everywhere. The cubic basis leads on asymmetric links, by $0.25$ over the
symmetric basis, and on the off-index scenarios; the symmetric basis leads on symmetric tails and on
rough misfit. Averaged over the six families the cubic basis leads narrowly, $0.452$ against $0.442$
and $0.425$, so it is the default, and an analyst who knows which departure they fear should choose
the basis from the row for that family rather than from the average.

\input{tab_families}

\subsection{The partition, Stukel's test and the calibration belt}\label{sec:verdicts}

\paragraph{Refining the partition} On clean data the default partition gives EDGE more
power than ten groups: it leads or ties the ten-group form in $155$ of the $158$ scenarios, with a mean
gain of $0.017$ in size-adjusted power. Against the partition tests the gain is larger, because the
Hosmer--Lemeshow statistic cannot use a finer partition without spending a degree of freedom on each
new group. Against the best of four partition tests (the Hosmer--Lemeshow statistic, its Farrington
correction, its equal-width form and the Pigeon--Heyse test), EDGE at the default
partition leads or ties in all $21$ scenarios that the specification selected as detectable and not
saturated; with ten groups for every test, in $17$ of the $19$ selected for that comparison.

\paragraph{Against Stukel's test} Stukel's score test is the established directed test for this
problem. The default basis stays within $0.02$ of Stukel's joint score in mean size-adjusted power in
every departure family, and over all $158$ scenarios it is $0.011$ behind (Monte Carlo $95\%$ interval
$0.010$ to $0.013$). On
the symmetric-tail family, the departure Stukel's one-parameter score was built for, the one-parameter
score is ahead of the default basis by $0.062$; there the symmetric basis is the one to use,
and in its score form it matches the one-parameter score, $0.002$ behind on average. In its unit form it
is level with Stukel's joint score on the same family, $0.005$ ahead. On clean data EDGE is
therefore comparable to Stukel's test; they differ in how they behave under corrupted records
(Section~\ref{sec:contamination}).

\paragraph{Against the calibration belt} On the symmetric-tail family the cubic basis leads the GiViTI
belt by $0.105$ on average, but not in every regime: in the scenarios with $12\%$ events the belt leads,
by $0.152$ and $0.124$ in the two probit scenarios and by smaller margins in four others. On the
asymmetric-link family the cubic basis is $0.075$ behind the belt. Both results have the same cause.
When events are scarce, or when the calibration curve bends once and smoothly, a polynomial fitted to
the whole risk scale uses information that a partition averages away. These are regimes in which the
belt, or for the single bend the cubic calibration test, is the better test, and
Section~\ref{sec:whennot} lists them as such.

\subsection{Against the smoothing and resampling tests}\label{sec:rivals}

Table~\ref{tab:rivals} gives raw rejection rates of le Cessie's smoothed test, the Liu projection test
and BAGofT beside EDGE on the $20$ link and tail scenarios at $n\ge500$. Le Cessie's test
holds its level in all ten matched null scenarios, between $0.040$ and $0.064$. On size-adjusted power
over the same $500$ replicates, EDGE leads or ties le Cessie's test in $19$ of the $20$
scenarios, is ahead after the Holm correction in $14$, and has mean power $0.451$ against $0.314$. It
leads the projection test in $17$ scenarios and BAGofT in $18$, by up to $0.364$ and $0.630$. All three
rivals lead it clearly in one scenario, the same one for all three: the crossover departure, in which
the calibration curve crosses the diagonal twice. There le Cessie's test rejects $496$ of $500$
replicates and the projection test $478$, against EDGE's $37$, and BAGofT rejects $155$ of
$200$ against $18$.

\input{tab_rivals}

The same three tests were run on the families they are built for, omitted terms and rough misfit:
le Cessie's test on the $56$ such scenarios with $n\le1000$, the projection test on the $19$ at
$n=500$ and BAGofT on six of those. Every matched null held its level, between $0.025$ and $0.070$.
Where the omitted term acts through the fitted risk, a quadratic term, EDGE leads or ties
le Cessie's test in $18$ of $21$ scenarios; where it does not, an interaction, le Cessie's test leads in
$19$ of $27$, and the projection test on all four continuous interactions. On the fastest
oscillation and the sawtooth BAGofT reaches $1.000$ and $0.805$ against EDGE's $0.295$ and $0.045$. A
basis of smooth shapes cannot follow a fast oscillation, and a test that learns its own partition can.

At $n=200$, where the default partition is ten groups and there is nothing to refine,
EDGE leads or ties le Cessie's test and the projection test in all six link and tail scenarios and
BAGofT in five. Its mean size-adjusted power is $0.160$ against $0.111$ and $0.125$, and $0.098$
against BAGofT's $0.073$ on the replicates BAGofT was run on. Two paired differences survive the Holm
correction, both in its favour and both on the asymmetric Stukel design, where its size-adjusted power
is $0.450$ against le Cessie's $0.330$ and $0.350$ against BAGofT's $0.185$.

\subsection{All scenarios, and what each test costs}\label{sec:census}\label{sec:compute}

\begin{figure}[htbp]\centering
\includegraphics[width=\textwidth]{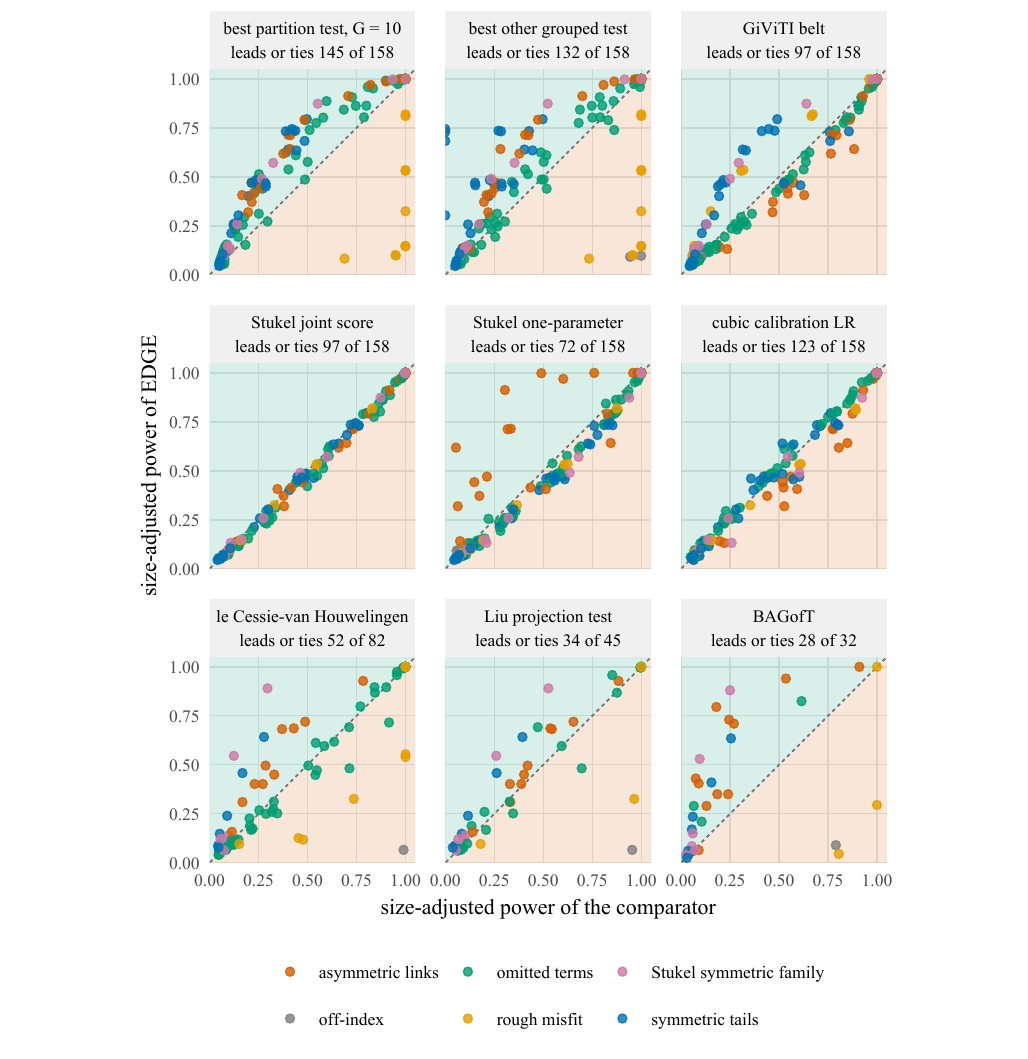}
\caption{EDGE at its default setting against every comparator, one point per alternative scenario. Both
axes are size-adjusted power, so each test is read at its own realised level; a point in the green
half is a scenario in which EDGE is ahead. The strip gives the number of scenarios in
which it leads or ties, within $0.01$. The first panel takes, in each scenario, the better of the
Hosmer--Lemeshow statistic and its Farrington correction at ten groups; the second, the best of the
equal-width Hosmer--Lemeshow statistic, the Pigeon--Heyse, Tsiatis and Xie tests. Le Cessie's test and
the two resampling tests were run on a subset of the scenarios, and the strip gives how many. Colour
is the departure family.}
\label{fig:scenarios}
\end{figure}

\begin{figure}[htbp]\centering
\includegraphics[width=\textwidth]{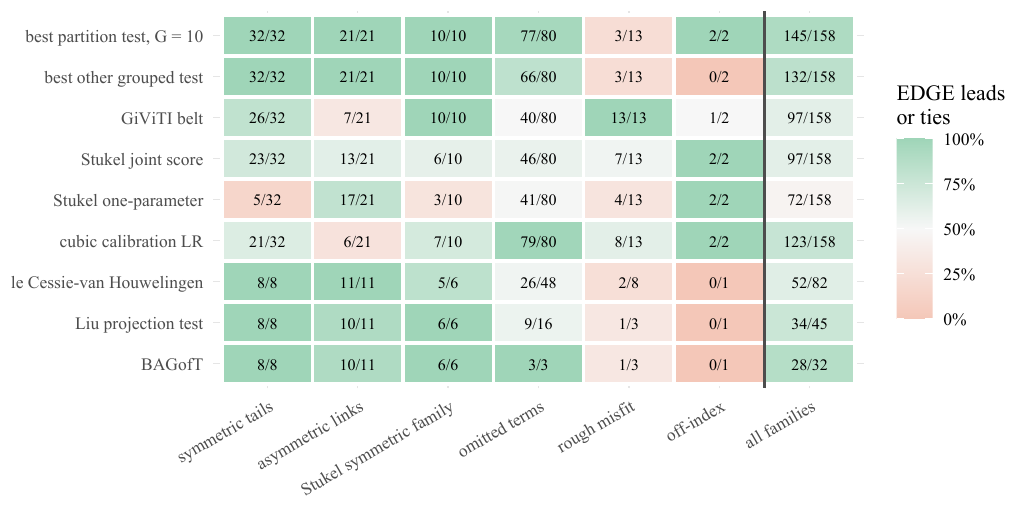}
\caption{The comparison of Figure~\ref{fig:scenarios} by departure family: each entry is the number of
scenarios in which EDGE leads or ties the comparator in that row. The last column is the
total over all families.}
\label{fig:census}
\end{figure}

Table~S7 of the Supporting Information gives, for each departure family, the mean difference in
size-adjusted power between EDGE at its default setting and each comparator, with a Monte Carlo $95\%$
interval. Over all $158$ scenarios EDGE is ahead of the Hosmer--Lemeshow statistic at ten groups by
$0.064$ ($0.063$ to $0.066$): by $0.18$ on asymmetric links and symmetric tails, $0.13$ on the Stukel
symmetric family and $0.07$ on omitted terms, and behind by $0.46$ on rough misfit, which a basis of
smooth shapes cannot follow. It is level with Stukel's joint score ($-0.011$), the one-parameter score
($-0.007$) and the cubic calibration test ($-0.008$), and $0.015$ ahead of the GiViTI belt, which leads by
$0.07$ on asymmetric links and trails by $0.10$ on symmetric tails. The tie with the cubic calibration
test is expected, because EDGE's default basis looks for the same shapes: a cubic in the predicted risk,
where that test fits a cubic in the logit. The two tests ask the same question of clean data; they
differ in how a record enters the answer, and that is what decides their behaviour when a few records
are wrong (Section~\ref{sec:whatprotects}). Figure~\ref{fig:scenarios} shows the
same comparison scenario by scenario and Figure~\ref{fig:census} by family; the counts in their strips
use a margin of $0.01$, within which two tests cannot be told apart. Against the other grouped tests
EDGE is behind in $14$ of the $80$ omitted-term scenarios, $10$ of the $13$ rough-misfit scenarios and
both off-index scenarios; $20$ of these $26$ losses are to Xie's and Tsiatis's tests, which group on the
covariates.

\begin{figure}[htbp]\centering
\includegraphics[width=\textwidth]{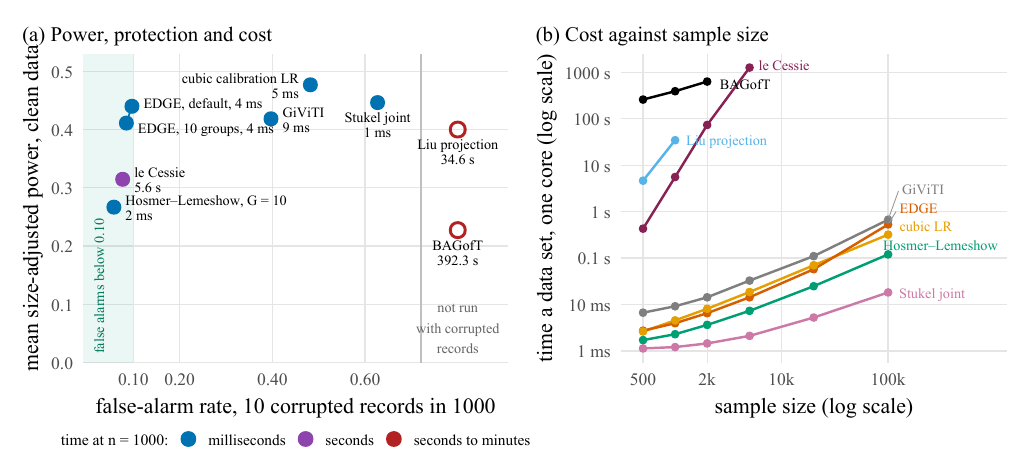}
\caption{Power, protection and cost of each test. (a)~Mean size-adjusted power on clean data
over the twenty scenarios on which all eight tests were run (link and tail departures and one
crossover), against the false-alarm rate when ten records in $1000$ carry a corrupted covariate and the
model is otherwise correct (corruption C1, Section~\ref{sec:contamination}); the shaded strip is below
$0.10$. Each label gives the median time for one data set at $n=1000$. The two resampling tests were
not run with corrupted records. The power axis ranks the tests on one family only, the one in which a
curve fitted to the whole risk scale is strongest; over all $158$ scenarios EDGE leads or
ties the cubic calibration test in $123$ (Figure~\ref{fig:scenarios}), as expected of two tests that
look for the same cubic shapes (Section~\ref{sec:census}). (b)~Median time against sample size, both axes
logarithmic, on one core. Each rival is timed as published: BAGofT from its own package at its
defaults, whose cost is $10{,}100$ model fits, le Cessie's test in its published $O(n^3)$ form, and the
projection test from our implementation at $B=250$. The tests without resampling are timed over eleven
repetitions, le Cessie's test over three at $n\le1000$; le Cessie's test at $n\ge2000$, the projection
test at $n=1000$ and BAGofT are timed once.
Colour gives the cost class in (a); EDGE is shown at the default partition and at ten groups.}
\label{fig:cost}
\end{figure}

Figure~\ref{fig:cost} sets power on clean data against protection and cost, on one core with nothing
else running. At $n=1000$ EDGE takes $4$ milliseconds, Stukel's joint score $1$, the
Hosmer--Lemeshow statistic $2$, the cubic calibration test $5$ and the GiViTI belt $9$; le Cessie's test
takes $5.6$ seconds, the projection test $35$ and BAGofT $392$, three to five orders of magnitude more.
On the twenty scenarios of panel (a) the cubic calibration test ($0.477$) is slightly ahead of EDGE
($0.440$) and Stukel's joint score ($0.446$) level with it, but neither survives one corrupted record in a thousand
(Section~\ref{sec:contamination}). Among the protected tests EDGE has the most power: at ten groups
$0.411$, against $0.267$ for the Hosmer--Lemeshow statistic, the only cheaper one, and $0.314$ for
le Cessie's test, at more than a thousand times the cost in its published form. These panel (a) values
use each test's full replicates; on the $500$ replicates of Table~\ref{tab:rivals} EDGE reads $0.451$
(Section~\ref{sec:rivals}).

The tests without resampling grow roughly in proportion to $n$: at $n=100{,}000$ EDGE takes half
a second, more than Stukel's score and the Hosmer--Lemeshow statistic ($18$ and $120$ milliseconds)
because it also builds the covariance of its partition. Le Cessie's test in its published form multiplies $n\times n$
matrices and grows with the cube of $n$: $74$ seconds at $n=2000$ and $21$ minutes at $n=5000$, where EDGE
takes $14$ milliseconds. The rates reported for le Cessie's test were computed with its implementation
in \texttt{ebrahim.gof}, which gives the same statistic. BAGofT's cost is fixed by its design rather than by $n$: $(1+\textit{nsim})\times
\textit{nsplits}$ model fits, ten minutes for one test at $n=2000$.

%% file: tables/tab_size.tex
\begin{table}[htbp]\centering\small
\caption{Size over the $131$ null scenarios of the study, for each basis and weighting at the
default partition. The three columns on the left summarise the realised size at the nominal $5\%$
level. The level check is made at both $5\%$ and $1\%$, and the last three columns
count the scenarios outside a band of three Monte Carlo standard errors at each level and at
either. No form passed in every scenario; under the fallback clause of the rule fixed in advance the
unit form, the specified default, was kept. Power is size-adjusted throughout.}
\label{tab:size}
\begin{tabular}{llrrrrrrr}\hline
 & & & \multicolumn{3}{c}{size at $5\%$} & \multicolumn{3}{c}{scenarios failing the gate} \\
\cmidrule(lr){4-6}\cmidrule(lr){7-9}
Basis & Weighting & Scenarios & Min & Median & Max & at $5\%$ & at $1\%$ & either \\ \hline
poly3 & score & 131 & 0.021 & 0.049 & 0.062 & 1 & 3 & 4 \\
poly3 & unit & 131 & 0.019 & 0.050 & 0.060 & 1 & 0 & 1 \\
stk & score & 131 & 0.027 & 0.049 & 0.064 & 1 & 9 & 10 \\
stk & unit & 131 & 0.013 & 0.048 & 0.059 & 0 & 3 & 3 \\
sym & score & 131 & 0.036 & 0.050 & 0.064 & 1 & 0 & 1 \\
sym & unit & 131 & 0.017 & 0.049 & 0.059 & 0 & 1 & 1 \\
\hline\end{tabular}\end{table}

%% file: tables/tab_families.tex
\begin{table}[htbp]\centering\small
\caption{Mean size-adjusted power of the three bases, unit weighting at the default partition, over
the six departure families. The last row is the macro-average of
the six family means, on which the cubic basis is the default.}
\label{tab:families}
\begin{tabular}{llrrrr}\hline
Family & Departure & Scenarios & EDGE-poly3 & EDGE-stk & EDGE-sym \\ \hline
1 & symmetric tails & 32 & 0.380 & 0.339 & 0.392 \\
2 & asymmetric links & 21 & 0.669 & 0.584 & 0.421 \\
3 & Stukel symmetric family & 10 & 0.464 & 0.451 & 0.481 \\
4 & omitted terms & 80 & 0.600 & 0.617 & 0.617 \\
5 & rough misfit & 13 & 0.507 & 0.572 & 0.576 \\
6 & off-index & 2 & 0.095 & 0.087 & 0.064 \\
\hline
\multicolumn{3}{l}{macro-average of the six} & 0.452 & 0.442 & 0.425 \\
\hline\end{tabular}\end{table}

%% file: tables/tab_rivals.tex
\begin{table}[htbp]\centering\small
\caption{EDGE against the three tests built without a partition, on the same data sets:
le Cessie's smoothed test \citep{le1991goodness}, the Liu projection test \citep{liu2024comprehensive} and
BAGofT \citep{BAGofT2019}, over the $20$ link and tail scenarios at $n\ge500$. Rejection rates at the
nominal $5\%$ level on replicates $1$--$500$ ($1$--$200$ for BAGofT); the highest rate in each row is in
bold, and $\star$ marks a paired difference from EDGE significant after the Holm correction.
EDGE is at its default partition. The closing rows summarise the table: after the Holm correction EDGE
is ahead of le Cessie's test and BAGofT in $14$ of the $20$ scenarios and of the projection test in
$7$, and each rival is ahead in one, the crossing curve; the mean is of raw rates, while the text and
Figure~\ref{fig:cost} give size-adjusted power. The rivals lead on the families they are built for,
interactions and rough misfit (Section~\ref{sec:rivals}, Table~\ref{tab:whennot}). Time is the median
for one data set at $n=1000$ on one core, le Cessie's test in its published form.}
\label{tab:rivals}
\begin{tabular}{llrrrrr}\hline
Departure & Design & $n$ & EDGE & le Cessie & Liu projection & BAGofT \\ \hline
\multicolumn{7}{l}{\emph{Symmetric tails}} \\
\quad cauchit link & high AUC & 460 & \textbf{0.470} & 0.200$^\star$ & 0.262$^\star$ & 0.140$^\star$ \\
\quad cauchit link & one covariate & 500 & \textbf{0.120} & 0.052$^\star$ & 0.082 & 0.055 \\
\quad cauchit link & high AUC & 760 & \textbf{0.654} & 0.316$^\star$ & 0.428$^\star$ & 0.245$^\star$ \\
\quad cauchit link & one covariate & 1000 & \textbf{0.254} & 0.094$^\star$ & 0.134$^\star$ & 0.080$^\star$ \\
\quad $t_4$ link & one covariate & 500 & 0.054 & 0.052 & 0.056 & \textbf{0.060} \\
\quad $t_4$ link & one covariate & 1000 & \textbf{0.080} & 0.056 & 0.060 & 0.035 \\
\multicolumn{7}{l}{\emph{Asymmetric links}} \\
\quad log--log link & high AUC & 380 & \textbf{0.356} & 0.184$^\star$ & 0.332 & 0.120$^\star$ \\
\quad log--log link & one covariate & 500 & \textbf{0.352} & 0.228$^\star$ & 0.318 & 0.090$^\star$ \\
\quad log--log link & base & 610 & \textbf{0.440} & 0.266$^\star$ & 0.412 & 0.140$^\star$ \\
\quad log--log link & high AUC & 640 & \textbf{0.606} & 0.348$^\star$ & 0.554 & 0.165$^\star$ \\
\quad log--log link & base & 1000 & \textbf{0.736} & 0.472$^\star$ & 0.642$^\star$ & 0.250$^\star$ \\
\quad log--log link & $12\%$ events & 1000 & 0.352 & 0.282$^\star$ & \textbf{0.358} & 0.095$^\star$ \\
\quad log--log link & one covariate & 1000 & \textbf{0.714} & 0.448$^\star$ & 0.560$^\star$ & 0.280$^\star$ \\
\quad Stukel, asymmetric & one covariate & 500 & \textbf{0.910} & 0.784$^\star$ & 0.876 & 0.535$^\star$ \\
\quad Stukel, asymmetric & one covariate & 1000 & \textbf{1.000} & 0.992 & 0.998 & 0.935$^\star$ \\
\multicolumn{7}{l}{\emph{Stukel symmetric family}} \\
\quad Stukel, long tails & one covariate & 500 & \textbf{0.096} & 0.068 & 0.088 & 0.055 \\
\quad Stukel, long tails & one covariate & 1000 & \textbf{0.148} & 0.102 & 0.112 & 0.070 \\
\quad Stukel, short tails & one covariate & 500 & \textbf{0.476} & 0.124$^\star$ & 0.248$^\star$ & 0.095$^\star$ \\
\quad Stukel, short tails & one covariate & 1000 & \textbf{0.900} & 0.320$^\star$ & 0.556$^\star$ & 0.270$^\star$ \\
\multicolumn{7}{l}{\emph{A crossing curve}} \\
\quad curve crossing the diagonal twice & one covariate & 1000 & 0.074 & \textbf{0.992}$^\star$ & 0.956$^\star$ & 0.775$^\star$ \\
\hline
\multicolumn{3}{l}{Mean rejection rate (raw)} & 0.440 & 0.319 & 0.402 & 0.224 \\
\multicolumn{3}{l}{EDGE ahead / behind after Holm} & -- & 14 / 1 & 7 / 1 & 14 / 1 \\
\multicolumn{3}{l}{Time a data set, $n=1000$} & 4 ms & 5.6 s & 35 s & 392 s \\
\hline\end{tabular}\end{table}

%% file: sections/06_corrupted.tex
\section{Corrupted records}\label{sec:contamination}

\subsection{False alarms from a handful of records}\label{sec:c1}

Table~S5 of the Supporting Information gives the false-alarm rate of every test under corruption C1: $k$ records whose
covariate has been multiplied by four, in a sample for which the model is otherwise correct. Over the
$24$ scenarios of C1 the slope on the corrupted covariate moves by less than $5\%$, the slope on the
second covariate by at most $1.9\%$ and the intercept by at most $0.02$ on the linear-predictor scale,
so the model under test is effectively the clean one; under the sign error the slope moves by up to
$14\%$ (Section~\ref{sec:mechanism}). The fitted risks of the clean records do not move either. Their
mean absolute distance from the true risk, an integrated calibration index \citep{austin2019ici}, is
$0.0185$ with one exaggerated record in $1000$ and $0.0199$ with ten, against $0.0189$ without
corruption, which is the estimation error of any fit at this size; the corrupted records
themselves are mispredicted by about $20$ percentage points. Ten sign errors move it to $0.027$. A test
that rejects here is reacting to a few records, not to the model that is used for everyone else.

One corrupted record in a thousand moves every record-level test. Stukel's joint score rises from the
nominal level to $0.121$, its one-parameter form to $0.115$, the cubic calibration test to $0.127$ and
the GiViTI belt to $0.089$. EDGE reads $0.056$ with the cubic basis and $0.058$ with the
symmetric basis. At five corrupted records the record-level tests are between $0.25$ and $0.40$ and 
EDGE is still at $0.057$. At ten, Stukel's joint score reaches $0.627$, rejecting a correct
model more often than not, while EDGE reads $0.097$. Every test holds its level in the
same design without corruption (the $k=0$ rows), so the difference is not that one test is
conservative.

The pooled tests are protected, as Section~\ref{sec:mechanism} predicts. Le Cessie's smoothed test
reads $0.055$, $0.052$, $0.052$ and $0.077$ at one, two, five and ten corrupted records: an isolated
record adds one term to a sum whose spread is set by all $n$ records. The Hosmer--Lemeshow statistic and
the Pigeon--Heyse test at ten groups are protected too: under the exaggeration they stay at or below
$0.058$ up to ten corrupted records in $1000$ and twenty-five in $5000$ (Table~\ref{tab:severity}),
although the Pigeon--Heyse test is conservative throughout, between $0.019$ and $0.035$ on clean data.
Other grouped tests are not protected. The equal-width Hosmer--Lemeshow statistic, whose extreme groups hold few
records, reaches $0.592$ at ten corrupted records in $1000$, and the two tests that group on the
covariates, whose outlying clusters can be small, reach $0.542$ (Tsiatis) and $0.606$ (Xie). In a smaller study of three corrupted
records in $500$ with the covariate multiplied by four or by eight, $100$ replicates a scenario, BAGofT
and EDGE with the cubic basis gave no sign of moving, while the projection test reached
$0.22$ at $\times8$ and EDGE with the symmetric basis $0.20$; with $100$ replicates a rise
of a few points could not be seen (Supporting Information S3).

Wrong outcomes are a different matter. When the ten highest-risk records in a
thousand have their event erased (C2), every test rejects more often, EDGE included, at
$0.667$ against Stukel's $0.718$ and the cubic test's $0.837$. Missing events at the top of the risk
scale are a miscalibration of the recorded data, and a calibration test should find them. A unit error
that makes predictions milder (C3) moves no test under the logistic truth, although under the two
misfit truths it costs some power, and duplicated records (C4) leave every test between $0.040$ and
$0.071$ (Supporting Information S2).

\subsection{What protects: pooling with unit weights}\label{sec:whatprotects}

Proposition~\ref{prop:bounded} contrasts a grouped statistic with an ungrouped direction that grows with
the linear predictor. A direction that is bounded, a polynomial in the predicted risk rather than in the
logit, might protect without grouping, and the last study of Section~\ref{sec:contamination-design}
tests this on the same data sets (Supporting Information S10). It does not. With one exaggerated record
in $1000$ the ungrouped score test for the cubic directions of EDGE rejects a correct model in $0.166$
of replicates and with ten in $0.641$, against $0.047$ and $0.101$ for EDGE at the default partition and
$0.051$ and $0.066$ at ten groups; with ten reversed records it rejects in $0.992$. EDGE with one record a
group behaves like Stukel's test ($0.148$ and $0.594$). Spiegelhalter's $z$ never rejects at $5\%$ on
development data, because its normal reference ignores the fit, but read at its own realised level it
rejects in $0.118$ and $0.546$. Grouping alone is not enough either: in the study of
Section~\ref{sec:c1}, EDGE's score form, which uses the same groups but divides by the information its
directions keep after the fit, rejects in $0.087$ and $0.463$ at ten groups with one and ten
exaggerated records, against $0.056$ and $0.085$ for the unit form. The weighted grouped tests of \citet{hosmer2002goodness}, which weight each
record within its group by a function of its covariates, reach $0.18$ and $0.27$ with ten exaggerated
records and $0.68$ and $0.82$ with ten reversed ones. On frozen predictions, where nothing is refitted, the picture changes, which locates the damage
(Supporting Information S10). The same ungrouped twin then reads $0.061$ and $0.096$ with one and ten
exaggerated records in $1000$, and Spiegelhalter's $z$ holds throughout ($0.090$ with ten reversed
records), as in Section~\ref{sec:realcorrupt}. Two things therefore make a test fragile: the refit,
whose effect a score test magnifies by dividing by the information left after it, and, without a refit,
directions that lean on the few most extreme records, since the twin on frozen predictions still reads
$0.311$ with ten reversed records, where EDGE at ten groups reads $0.082$. EDGE's unit form avoids both:
its divisor is the variance of the clean groupmates, which neither the record nor the refit can make
small. The same holds with twelve
covariates at $n=2000$: with ten reversed records in one covariate EDGE reads $0.091$ at the default
partition and $0.048$ at ten groups, against $0.315$ for Stukel's test and $0.292$ for the ungrouped twin.

A routine alternative is to find the corrupted records with influence diagnostics and test on the rest.
Removing every record whose $\Delta X^2$ exceeds $4$ or whose $\Delta\beta$ exceeds $1$
\citep{pregibon1981logistic,hosmer2013applied} removed about $3.7\%$ of the records of a clean data set and
made every test reject a correct model in more than $0.95$ of replicates; removing only the records
with $\Delta\beta>1$ removed none, corrupted or not, and changed nothing. At an extreme prediction a
corrupted record's leverage and residual are both small on the fitted scale, so the diagnostics do not
single it out.

\subsection{The partition, not the contamination rate}\label{sec:partition}

On the same data the Hosmer--Lemeshow statistic reads $0.437$ at fifty corrupted records in $5000$ at
the default partition, where groups hold twenty-five records, and $0.088$ at ten groups, where they
hold five hundred; EDGE at the default partition reads $0.740$, and the record-level
tests between $0.87$ and $0.98$.

\begin{figure}[htbp]\centering
\includegraphics[width=\textwidth]{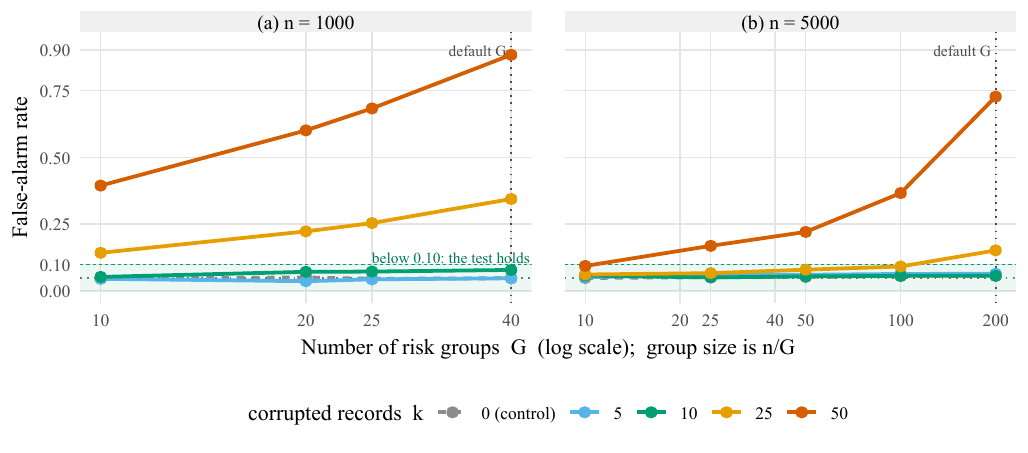}
\caption{False-alarm rate of EDGE (cubic basis, unit form) under a correct model when $k$
records carry a corrupted covariate, against the number of risk groups. Each line is a fixed $k$; the
shaded area lies below $0.10$, where the test is read as holding, the dotted vertical line marks the
default partition, and the dashed grey line is the uncorrupted control. (a)~$n=1000$; (b)~$n=5000$. The
same contamination rate, $0.5\%$, gives different false-alarm rates in (a) and (b): the size of the
group, not the rate, decides.}
\label{fig:partition}
\end{figure}

A second study varied the number of groups with the data held fixed, so that the same replicate is
tested at every partition (Figure~\ref{fig:partition}). The contamination \emph{rate} does not decide
the outcome: twenty-five corrupted records in $5000$ and five in $1000$ are both $0.5\%$, and 
EDGE reads $0.152$ and $0.048$, whereas the partition changes it: at $n=5000$ with twenty-five corrupted
records the same data give $0.062$ at ten groups and $0.152$ at two hundred; with fifty they give
$0.094$ and $0.726$. Without corruption the rate stays between $0.046$ and $0.057$ at every
partition.

The reason is in \eqref{eq:twomoments}. Corrupted covariates produce the most extreme predictions, so
they fill the top and bottom groups first; about $0.35k$ to $0.38k$ of them reach the top group at the
default partition, and about $0.40k$ at ten groups. The shift in that group's residual grows as
$\theta m_c/\sqrt{(m-m_c)\bar v}$, which rises steeply as $m_c$ approaches $m$. At the default partition
the group size $m$ stays near twenty-five, so the \emph{count} of corrupted records decides: in the
partition study ten in $1000$ put an average of $3.7$ into the top group and the test reads $0.079$,
and fifty in $5000$ put $17.8$ there, leaving about seven clean members, and it reads $0.726$. At ten
groups $m=n/10$ grows with the sample; the same data put $4.0$ and $19.9$ records into top groups of
$100$ and $500$, and the test reads $0.053$ and $0.094$. The partition study drew its own data sets, so
its rates differ from those of Table~S5 by Monte Carlo error.

\subsection{The severity of the error}\label{sec:severity}

A sign error (C1$'$) reverses the prediction instead of exaggerating it: a high-risk patient is placed
at the bottom of the risk scale with the outcome intact, and $\theta$ in \eqref{eq:twomoments} rises
from about $0.3$ to about $0.7$. The fit also moves more, by $14\%$ in the slope, and the two channels
were not separated. Table~\ref{tab:severity} gives both errors. With the cubic basis at
the default partition EDGE holds to about two corrupted records in $1000$ and five in
$5000$ under the sign error, against about ten under the exaggeration: at ten records in $1000$ it
reads $0.381$. Two settings do much better. With ten groups the cubic basis holds to five records in
$1000$ and ten in $5000$. The symmetric basis at the default partition holds to ten in both, reading
$0.098$ and $0.058$, although against many exaggerated records it is no better than the cubic basis
($0.745$ at fifty in $5000$). The symmetric basis at ten groups is the exception to the partition
lever: under the sign error it is less protected than at the default partition ($0.431$ against
$0.098$ at ten records in $1000$), which we observe but cannot explain from \eqref{eq:twomoments}.
Whatever the setting, the ordering of
the tests is unchanged. At one corrupted record in a thousand EDGE reads $0.054$ against
Stukel's $0.364$, the cubic test's $0.389$ and the belt's $0.268$; at five, $0.124$ against $0.89$,
$0.89$ and $0.80$.

\input{tab_severity}

\subsection{Protection when records may be wrong}\label{sec:edgefr}

For the cubic basis the partition sets a trade-off between power and protection, and it is chosen
before the data are seen. The default partition has more power: on clean data it leads ten groups by
$0.017$ on average over the $158$ scenarios ($0.016$ to $0.018$), and by much more when the misfit is
confined to the extreme risks (Supporting Information S10). With the true risk of the top $2\%$ of
patients $40\%$ below its prediction its size-adjusted power is $0.42$ at $n=1000$, against $0.19$ at ten
groups; with that risk $20\%$ below its prediction at $n=5000$, $0.62$ against $0.29$; and with the ten
highest-risk outcomes erased (C2), $0.60$ against $0.26$. Ten groups tolerate more corrupted records: they read $0.085$ with
ten exaggerated records in $1000$ and $0.107$ with fifty in $5000$, where the default partition reads
$0.097$ and $0.740$, and under the sign error $0.081$ with five records in $1000$ and $0.077$ with ten in
$5000$, where the default partition holds to about two in $1000$. The partition is therefore chosen in the
analysis plan, from what is known about how the data were collected, and not after the results are
seen: running both and reporting the more convenient would bring back the dependence on $G$ that
Section~\ref{sec:realpartition} shows for the Hosmer--Lemeshow test. The software reports both; if only
the default partition rejects, the signal lies in the extreme groups, and their records should be
checked.

At ten groups the plain Hosmer--Lemeshow statistic holds its level further than EDGE
once there are more than a few reversed records: $0.099$ against $0.182$ at ten in $1000$, and $0.126$
against $0.249$ at twenty-five in $5000$. Up to five reversed records in $1000$ both stay below $0.10$
($0.081$ and $0.056$), although the Hosmer--Lemeshow statistic already rejects less often on the same
data sets (exact McNemar $p=0.013$).
The Pigeon--Heyse test holds further still, but it is conservative on clean data. What EDGE
offers at ten groups is power, $0.411$ against $0.267$ (Section~\ref{sec:census}).

Pooling only the two extreme groups and keeping the default partition in the middle matched ten groups
up to about ten corrupted records in $1000$ and fifty in $5000$, was less protected beyond, and had the
same power; removing the extreme records instead removes the signal with them (Supporting Information
S5).

Section~\ref{sec:whentouse} turns these results into a rule.

\subsection{The cost in power, and robust alternatives}\label{sec:cost}\label{sec:robustest}

Under corruption a rejection can come from the misfit or from the corrupted records. The raw rejection
rate barely moves within the protected range: against a complementary log--log truth at $n=1000$ EDGE
rejects in $0.818$ of replicates on clean data, $0.785$ with one corrupted record and $0.802$ with ten.
Part of that is false alarm, so power is also read at the critical value of the test's own
\emph{contaminated} null, the logistic truth at the same sample size with the same records corrupted,
which the analyst cannot know but which isolates the misfit. Read this way EDGE falls from $0.826$
to $0.771$ at one corrupted record and $0.648$ at ten; against a probit truth at $n=5000$ it falls
from $0.278$ to $0.233$ at five and $0.207$ at ten. Corrupted records and a real tail misfit occupy the
same extreme groups, so the averaging that absorbs the first also weakens the second. Neither basis pays
consistently less: read at the contaminated null, at five corrupted records in $5000$ against the
probit truth the symmetric basis loses nothing and the cubic basis $0.045$, while against the
complementary log--log truth at one and two records in $1000$ the symmetric basis loses more ($0.084$
against $0.055$; $0.153$ against $0.096$). Beyond the protected range the test has
no power left: at twenty-five corrupted records in $5000$ its raw rejection rate against the probit
truth, $0.143$, is below its false-alarm rate on a correct model, $0.151$, and read at its
contaminated null it is $0.045$.

A natural objection is that corrupted records call for a robust fit. We fitted every contaminated data
set both by maximum likelihood and by the Huber-type quasi-likelihood estimator of
\citet{cantoni2001robust}, which bounds the Pearson residual (Huber's $\psi$ at $1.345$) but puts no
weight on the covariates, and applied every test to both fits. After a robust fit, five of ten tests no longer hold their level on clean data: both Stukel
score tests ($0.199$ and $0.223$) and three of EDGE's four forms, the symmetric basis in
its unit form ($0.164$) and both score forms. Of the five that do, none has its false alarms reduced: at
ten corrupted records the GiViTI belt reads $0.379$ after maximum likelihood and $0.400$ after the
robust fit, the cubic calibration test $0.465$ and $0.466$ (on this study's own data sets), and EDGE $0.089$ and $0.087$.
A corrupted record's prediction is formed at its corrupted covariate, so refitting does not reach it;
restoring the slope, if anything, makes that prediction more extreme. Refitting the model robustly is therefore not a substitute for a test
whose own statistic bounds a record's influence \citep{hampel1986robust,heritier1994robust}
(Supporting Information S4).

The fair robust competitor is therefore a robust \emph{test}. We compared EDGE with the
robust quasi-deviance test of \citet{cantoni2001robust} for Stukel's two constructed covariates, on
$1000$ fresh data sets a scenario at $n=1000$, in two forms: a Huber form, which bounds the Pearson
residual, and a Mallows form, which also down-weights records of high leverage and is the form meant
for wrong covariates. Both hold their level on clean data ($0.049$ and $0.047$). The Mallows form holds
below $0.10$ with one or two corrupted records, although with two reversed records it already rejects
more often than EDGE on the same data sets ($0.075$ against $0.056$, exact McNemar
$p=0.004$), and it does not hold beyond: with ten
exaggerated records it reads $0.168$ and with ten reversed ones $0.789$, where EDGE at ten
groups reads $0.063$ and $0.192$ on the same data sets (exact McNemar $p<10^{-20}$ in both). The Huber
form is less protected still ($0.205$ and $0.923$). On clean data the robust tests are also less
powerful: against a complementary log--log truth their size-adjusted power is $0.610$, against 
EDGE's $0.779$ at ten groups and $0.795$ at the default partition. Robustifying the
record-level test protects it against one or two exaggerated records and one reversed one, at a cost
in power; pooling protects further
at no such cost.

%% file: tables/tab_severity.tex
\begin{table}[htbp]\centering\small
\caption{False-alarm rate at the nominal $5\%$ level under a correct logistic model when $k$ records
carry a corrupted covariate, for the two errors of Section~\ref{sec:contamination-design}: the
covariate multiplied by four (exaggeration) and by $-4$ (sign error). EDGE is shown with the cubic
and symmetric bases at the default partition and at ten groups. HL-ew is the Hosmer--Lemeshow statistic with equal-width risk groups; PH is the
Pigeon--Heyse test and Tsiatis the score test on indicators of ten $K$-means clusters in covariate space, all at ten groups.
$1000$ replicates a scenario, every test on the same data sets; the Monte Carlo standard error of
a rate of $0.05$ is $0.007$. A replicate with no $p$-value counts as a non-rejection. Shaded cells exceed
$0.10$, the rate above which a test is read as no longer holding its level.}
\label{tab:severity}
\begin{adjustbox}{max width=\textwidth}
\begin{tabular}{rrrrrrrrrrr}\hline
 & & \multicolumn{4}{c}{EDGE} & \multicolumn{4}{c}{grouped rivals} & \\
\cmidrule(lr){3-6}\cmidrule(lr){7-10}
$n$ & $k$ & cubic, default & cubic, $G{=}10$ & symmetric, default & symmetric, $G{=}10$ & HL, $G{=}10$ & HL-ew & PH & Tsiatis & Stukel joint \\ \hline
\multicolumn{11}{l}{\emph{Exaggeration}, $x\mapsto4x$} \\
1000 & 0 & 0.045 & 0.040 & 0.044 & 0.047 & 0.040 & 0.040 & 0.019 & 0.059 & 0.044 \\
1000 & 1 & 0.056 & 0.056 & 0.058 & 0.056 & 0.051 & \cellcolor{black!12}0.151 & 0.033 & 0.077 & \cellcolor{black!12}0.121 \\
1000 & 2 & 0.059 & 0.053 & 0.067 & 0.054 & 0.044 & \cellcolor{black!12}0.246 & 0.023 & \cellcolor{black!12}0.125 & \cellcolor{black!12}0.217 \\
1000 & 5 & 0.057 & 0.051 & 0.048 & 0.067 & 0.042 & \cellcolor{black!12}0.444 & 0.027 & \cellcolor{black!12}0.287 & \cellcolor{black!12}0.398 \\
1000 & 10 & 0.097 & 0.085 & 0.059 & \cellcolor{black!12}0.114 & 0.058 & \cellcolor{black!12}0.592 & 0.037 & \cellcolor{black!12}0.542 & \cellcolor{black!12}0.627 \\
5000 & 0 & 0.048 & 0.051 & 0.053 & 0.053 & 0.055 & 0.047 & 0.032 & 0.037 & 0.047 \\
5000 & 5 & 0.050 & 0.048 & 0.045 & 0.049 & 0.037 & \cellcolor{black!12}0.380 & 0.020 & \cellcolor{black!12}0.173 & \cellcolor{black!12}0.252 \\
5000 & 10 & 0.060 & 0.047 & 0.052 & 0.056 & 0.043 & \cellcolor{black!12}0.586 & 0.022 & \cellcolor{black!12}0.325 & \cellcolor{black!12}0.472 \\
5000 & 25 & \cellcolor{black!12}0.151 & 0.065 & 0.099 & 0.090 & 0.050 & \cellcolor{black!12}0.789 & 0.038 & \cellcolor{black!12}0.680 & \cellcolor{black!12}0.801 \\
5000 & 50 & \cellcolor{black!12}0.740 & \cellcolor{black!12}0.107 & \cellcolor{black!12}0.745 & \cellcolor{black!12}0.201 & 0.088 & \cellcolor{black!12}0.958 & 0.052 & \cellcolor{black!12}0.907 & \cellcolor{black!12}0.980 \\
\hline
\multicolumn{11}{l}{\emph{Sign error}, $x\mapsto-4x$} \\
1000 & 0 & 0.051 & 0.040 & 0.047 & 0.051 & 0.052 & 0.052 & 0.032 & 0.054 & 0.048 \\
1000 & 1 & 0.054 & 0.050 & 0.056 & 0.069 & 0.057 & \cellcolor{black!12}0.392 & 0.032 & \cellcolor{black!12}0.179 & \cellcolor{black!12}0.364 \\
1000 & 2 & 0.072 & 0.072 & 0.079 & 0.095 & 0.058 & \cellcolor{black!12}0.614 & 0.030 & \cellcolor{black!12}0.356 & \cellcolor{black!12}0.584 \\
1000 & 5 & \cellcolor{black!12}0.124 & 0.081 & 0.095 & \cellcolor{black!12}0.225 & 0.056 & \cellcolor{black!12}0.892 & 0.040 & \cellcolor{black!12}0.714 & \cellcolor{black!12}0.893 \\
1000 & 10 & \cellcolor{black!12}0.381 & \cellcolor{black!12}0.182 & 0.098 & \cellcolor{black!12}0.431 & 0.099 & \cellcolor{black!12}0.988 & 0.064 & \cellcolor{black!12}0.949 & \cellcolor{black!12}0.992 \\
5000 & 0 & 0.060 & 0.058 & 0.050 & 0.054 & 0.049 & 0.047 & 0.035 & 0.044 & 0.058 \\
5000 & 5 & 0.072 & 0.061 & 0.070 & 0.080 & 0.051 & \cellcolor{black!12}0.776 & 0.034 & \cellcolor{black!12}0.443 & \cellcolor{black!12}0.746 \\
5000 & 10 & \cellcolor{black!12}0.139 & 0.077 & 0.058 & \cellcolor{black!12}0.163 & 0.054 & \cellcolor{black!12}0.972 & 0.039 & \cellcolor{black!12}0.721 & \cellcolor{black!12}0.974 \\
5000 & 25 & \cellcolor{black!12}0.794 & \cellcolor{black!12}0.249 & \cellcolor{black!12}0.291 & \cellcolor{black!12}0.543 & \cellcolor{black!12}0.126 & \cellcolor{black!12}1.000 & 0.080 & \cellcolor{black!12}0.998 & \cellcolor{black!12}1.000 \\
5000 & 50 & \cellcolor{black!12}1.000 & \cellcolor{black!12}0.903 & \cellcolor{black!12}0.999 & \cellcolor{black!12}0.947 & \cellcolor{black!12}0.575 & \cellcolor{black!12}1.000 & \cellcolor{black!12}0.493 & \cellcolor{black!12}1.000 & \cellcolor{black!12}1.000 \\
\hline\end{tabular}\end{adjustbox}\end{table}

%% file: sections/07_realdata.tex
\section{A clinical cohort}\label{sec:realdata}

The Diabetes 130-US hospitals cohort \citep{strack2014impact,clore2014uci} records $101{,}766$
admissions of patients with diabetes at $130$ hospitals over ten years, with readmission within
thirty days as the outcome. Patients recur across admissions, so we follow the data set's authors and
keep one admission per patient, the first, excluding admissions that ended in death or discharge to
hospice, whose patients cannot be readmitted, and the three records whose sex is unknown. That leaves
$69{,}987$ patients. We split them in half at random, fitted a logistic model on the first half, and
froze its predictions for the second. The model uses the routinely recorded predictors: age, sex, length
of stay, the numbers of laboratory procedures, other procedures, medications, outpatient, emergency and
inpatient visits in the preceding year and diagnoses, insulin use in four levels, a change of diabetes
medication, and any diabetes medication. No attempt was made to
make the model fit well or badly; it discriminates weakly, with an area under the curve of $0.60$ on
the validation half.

\subsection{The partition on real data}\label{sec:realpartition}

On the development half ($34{,}993$ patients) the fit is checked as the partition is refined from
$G=10$ to $G=2000$ (Figure~\ref{fig:diabetes}(b)). EDGE rejects at every partition,
with $p$ between $4\times10^{-5}$ and $7\times10^{-4}$. The Hosmer--Lemeshow statistic rejects at ten
groups ($p=0.041$) and at no finer partition, with $p$ between $0.10$ and $0.50$ from twenty groups on;
its Farrington correction behaves the same way. Two analysts who chose $G=10$ and $G=50$ would reach
opposite conclusions about the same model \citep{paul2013power}; EDGE's reference does not change
with the number of groups (Theorem~\ref{thm:growingG}).

\subsection{Validation}\label{sec:realvalidation}

On the validation half ($34{,}994$ patients) nothing is refitted, so $\Omega=I$ exactly, and because no
score equation absorbs the intercept, the statistic carries it as a fourth column beside the default
basis. The usual check, the calibration intercept-and-slope test of \citet{cox1958} and
\citet{Miller1991}, rejects: the calibration slope is $0.85$, so the predictions are too spread out, and
the likelihood-ratio statistic is $13.3$ on two degrees of freedom, $p=0.001$. The miscalibration is
small in absolute terms: the integrated calibration index of \citet{austin2019ici} is $0.007$, with E50
$0.007$ and E90 $0.008$, so the predictions are off by less than one percentage point for nine patients in
ten; with $35{,}000$ patients the tests detect it, and whether it matters clinically is a separate
question \citep{nattino2020assessing,vancalster2019}. EDGE
rejects at every partition, with $S$ between $23$ and $41$ on four degrees of freedom and $p$ between
$1\times10^{-4}$ and $3\times10^{-8}$, and the GiViTI belt in its external mode rejects as well
($p=1.5\times10^{-8}$). The Hosmer--Lemeshow statistic rejects at five of the eight partitions, with $p$
from $5\times10^{-6}$ at ten groups to $0.062$, $0.15$ and $0.055$ at five hundred, a thousand and two
thousand groups.

\begin{figure}[htbp]\centering
\includegraphics[width=\textwidth]{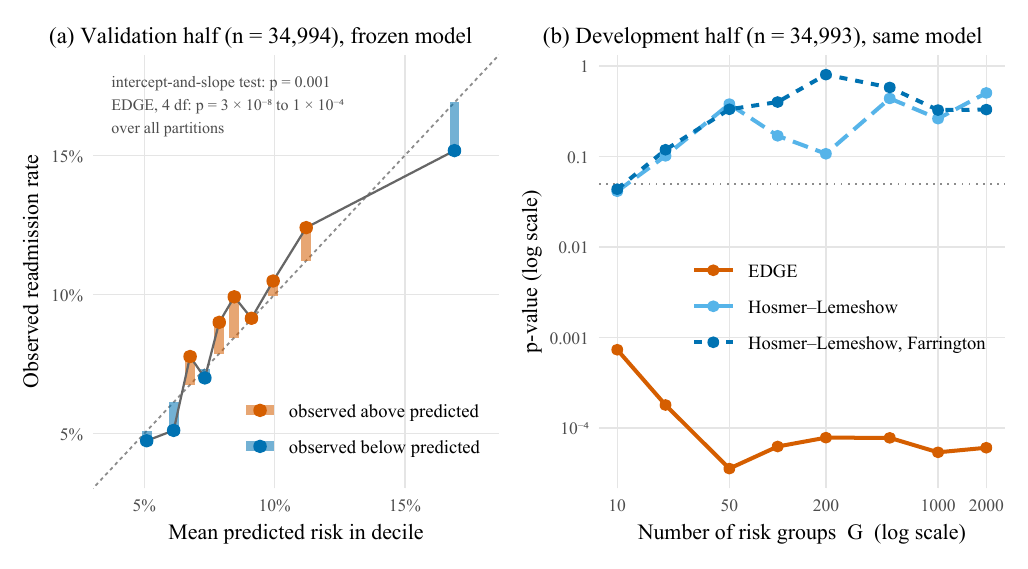}
\caption{The Diabetes-130 cohort, one admission per patient. (a)~Validation half
($n=34{,}994$), frozen model; observed readmission rate against mean predicted risk in ten
equal-frequency risk groups, with the gap from the diagonal drawn as a bar. (b)~Development
half ($n=34{,}993$), the same model; the $p$-value of EDGE, the Hosmer--Lemeshow statistic
and its Farrington correction as the partition is refined from $G=10$ to $G=2000$, on a
logarithmic scale; the dotted line is $0.05$.}
\label{fig:diabetes}
\end{figure}

The shape of the miscalibration is not a simple tilt. Figure~\ref{fig:diabetes}(a) shows the
top decile and the second over-predicted and the middle deciles under-predicted, by $1.0$ to $1.5$
percentage points in deciles three, five, six and nine. A grouped test restricted to a tilt, the intercept and slope columns alone, does
not reject at any partition ($p$ between $0.08$ and $0.34$): at ten groups only about $2$ of the
statistic's $23$ lie in the intercept and slope directions, and the rest in the curvature. The grouped
tilt is less sensitive than the record-level Cox--Miller test to a calibration slope of $0.85$, so 
EDGE complements the calibration intercept and slope and does not replace them.

\begin{figure}[htbp]\centering
\includegraphics[width=\textwidth]{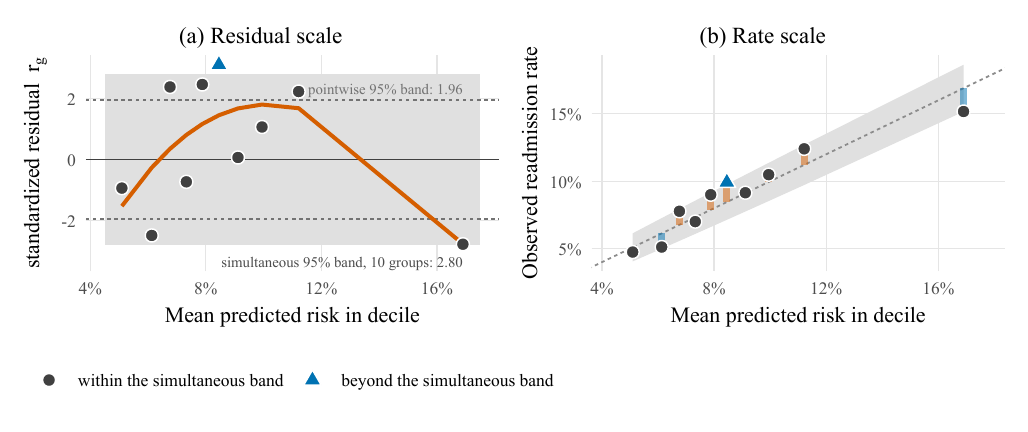}
\caption{The residual band implied by EDGE, on the validation half ($n=34{,}994$, frozen
model, ten equal-frequency risk groups). (a)~The standardized residual
$r_g=(O_g-E_g)/\sqrt{V_g}$ of each decile. Nothing was estimated from these outcomes, so each residual
is standard normal under a calibrated model; the shaded band is the simultaneous $95\%$ band over all
ten groups, a Bonferroni band for ten independent residuals, which is valid because nothing was refitted, and the dashed lines the pointwise band. Triangles mark the deciles beyond the simultaneous
band. The curve is the projection of $r$ onto the test's basis, whose squared length is the statistic.
(b)~The same band on the readmission-rate scale.}
\label{fig:belt}
\end{figure}

Figure~\ref{fig:belt} draws the ten standardized residuals against the reference the test uses, with
the pointwise band $\pm1.96$ and the simultaneous band $\pm2.80$ over all ten groups. Examining observed
and expected events group by group is standard practice after a Hosmer--Lemeshow test
\citep{hosmer2013applied}; what the figure adds is that the curve drawn through the residuals is the projection $P_Z r$, whose squared length is the statistic, so
the verdict and its location are read from the same picture. Six deciles clear the pointwise band and
one clears the simultaneous one, the sixth, at $r_g=3.1$; the top decile, at $-2.79$, sits on its edge.

For a model used at a threshold the shape has a direct cost. The highest-risk tenth of patients are
predicted to be readmitted at $16.9\%$ and are readmitted at $15.2\%$, $1.7$ percentage points fewer
than predicted, while patients predicted at $7$ to $11\%$ carry a risk up to $1.5$ points higher
than reported. An intercept and slope, the ``weak'' level of the calibration hierarchy
\citep{vancalster2016hierarchy}, record that the model is miscalibrated but not where. A smoothed calibration curve
\citep{vancalster2019,austin2019ici} shows where too, and it remains the first display; EDGE adds a
test whose verdict does not depend on the smoother or the partition, and whose residuals are read on
the same picture.

\subsection{Corrupted records in this cohort}\label{sec:realcorrupt}

The experiment of Section~\ref{sec:contamination} can be repeated on this cohort. On the real
covariates and the frozen predictions we draw the outcome from the model itself, so that the model is
correct by construction and every rejection is a false alarm. Then $k$ records have one covariate
multiplied by four, with their prediction recomputed at the corrupted value and their outcome kept.
There are $500$ replicates at each of $k=0,10,50,100$ in the validation half; the Monte Carlo standard
error of a rate of $0.05$ is $0.010$.

Corrupting the number of laboratory procedures moves no test at any $k$: at a hundred corrupted
records EDGE reads $0.036$, Stukel's joint score $0.052$ and the cubic calibration test
$0.048$. Its coefficient is small, so a fourfold error moves a record's linear predictor by $0.23$ on
average, less than the spread of the linear predictor across the cohort ($0.37$), and no record becomes
extreme.

The number of previous inpatient admissions is the model's largest coefficient, $0.38$. Only $12\%$ of
patients have any previous admission, so a fourfold error changes about one corrupted record in eight,
but it moves that record's linear predictor by $1.7$ on average, more than four times the cohort's
spread. This is an error that passes a range check, because four previous admissions is a plausible
value. It is also the covariate on which the experiment could show most, and it was chosen for that
reason after the first showed nothing. Of a hundred corrupted records about twelve therefore change, and the separation of
Section~\ref{sec:contamination} appears (Table~\ref{tab:cohortcorrupt}).

\begin{table}[htbp]\centering\small
\caption{False-alarm rate on the cohort's validation half ($n=34{,}994$) when $k$ records have the
number of previous inpatient admissions multiplied by four and the model is otherwise correct by
construction. The grouped tests use frozen predictions in the external mode (four columns); the
default partition has $1400$ groups. $500$ replicates at each $k$.}
\label{tab:cohortcorrupt}
\begin{tabular}{rrrrrr}\hline
$k$ & EDGE (default $G$) & EDGE ($G{=}10$) & HL ($G{=}10$) & Stukel joint & cubic LR \\ \hline
0   & 0.050 & 0.044 & 0.048 & 0.050 & 0.042 \\
10  & 0.042 & 0.042 & 0.050 & 0.102 & 0.080 \\
50  & 0.086 & 0.060 & 0.060 & 0.254 & 0.198 \\
100 & 0.080 & 0.054 & 0.048 & 0.448 & 0.314 \\ \hline
\end{tabular}
\end{table}

About twelve changed records among $34{,}994$ patients raise Stukel's score test to $0.448$ and the
cubic calibration test to $0.314$. At the default partition EDGE rises to $0.086$ with
fifty corrupted records, about six of them changed, and stays below $0.09$; at ten groups it stays
within Monte Carlo error of the nominal level.

A simulation of external validation in one design with five covariates (Supporting Information S7) sets
EDGE against the usual tests there. At ten groups its mean size-adjusted power over ten departures is
$0.415$, against $0.420$ for the GiViTI belt and $0.410$ for Cox's recalibration test, which in most data
sets return the same $p$-value; the two lead by up to $0.14$ on a shift and a threshold term, and EDGE by
up to $0.21$ on a U-shaped term and an interaction. With ten reversed records in $1000$ EDGE reads
$0.082$, below the bar of $0.10$, where Cox's test reads $0.158$ and the belt $0.347$; the
Hosmer--Lemeshow statistic and Spiegelhalter's $z$ stay below it too, with less power ($0.335$ and $0.271$).

\subsection{Monitoring a deployed model}\label{sec:stream}

Once the risk groups are fixed by cut points chosen in advance from reference predictions, the
external-mode statistic depends on the data only through four sums per group: observed events, expected
events, the binomial variance and the count. It can therefore be kept up to date as patients arrive:
each new patient updates one group, and the test is recomputed from the $G$ group summaries without
revisiting earlier records. The streamed statistic equals the external statistic computed in one pass on
all the records with the same cut points, whatever the order and batching of the updates; these fixed
groups drift away from equal size when the risk mix of the monitored patients changes. Provided every
group keeps receiving patients, it keeps its $\chi^2_{d+1}$ reference under a calibrated model
(Supporting Information S8). In a simulation fixed before it was run, at ten groups and with $2000$
replicates, the level was $0.047$ and $0.054$ after $10{,}000$ and $100{,}000$ patients, and $0.050$ and
$0.052$ when every monitored patient came from a risk distribution different from the reference; it
detected an overfitted model, calibration slope $0.8$, in $0.99$ of replicates after $2000$ patients.
The statistic accumulates since deployment, so it answers whether the model has been calibrated over
that whole period; to follow drift, the summaries are restarted each period, a month or a quarter, and
each period is tested on its own.
Adding a hundred patients takes about half a millisecond and the test about $1.5$ milliseconds on one
core, however many patients have been seen. One test at any time is valid, but acting on the first
rejection over repeated looks is not: ten looks at $5\%$ raised the false-alarm rate to $0.21$, and
splitting the level over ten looks fixed in advance, $0.5\%$ each, brought it to $0.030$. In
\texttt{ebrahim.gof} the monitor is \texttt{edge.stream()}.

%% file: sections/08_discussion.tex
\section{Discussion}\label{sec:discuss}

\subsection{What grouping still buys}\label{sec:asymmetry}

Grouping was introduced so that a goodness-of-fit test for logistic regression would have a reference
distribution. It also bounds the influence of a record whose prediction is wrong, because the record's
residual is divided by the variance of its clean groupmates. The tests that left grouping behind to
gain power lost that bound: the record-level tests lose their level with one corrupted record in a
thousand (Section~\ref{sec:c1}). So do grouped tests whose extreme groups are small or that group on the
covariates, and so do tests along bounded directions once the model is refitted
(Section~\ref{sec:whatprotects}).
Among the pooled tests compared here, EDGE is the most powerful on link and tail departures, and the Hosmer--Lemeshow statistic
at ten groups holds its level further under reversed records (Section~\ref{sec:edgefr}).

On clean data EDGE is comparable to Stukel's score test and the cubic calibration test.
Its advantage is that it keeps its level when a few records are wrong, at a modest cost in
power. The protection
has a size, which depends on the error and the partition (Section~\ref{sec:edgefr}), and it does not
extend to wrong outcomes, which a calibration test should detect. A grouped test that rejects while
covariate errors are suspected is therefore reporting something other than those errors; the same
test rejecting while outcome errors are suspected may be reporting them, and the data should be checked before the model is rebuilt. Tolerating a few
wrong inputs is the right behaviour when the question is whether the model is right, at development
and validation. Once the model is deployed, a patient whose input is wrong receives a wrong
prediction whatever the model's calibration; finding that patient is the job of checks at data
entry, not of a calibration test.

\subsection{Recommended use}\label{sec:whentouse}

\begin{enumerate}
\item Use the cubic basis in its unit form at the default partition, $G=\max(10,\lceil n/25\rceil)$, and
      report the calibration intercept and slope beside it; \texttt{edge.gof(fit)} reports it beside
      ten groups.
\item If the data are known to carry entry errors in the predictors, as routinely collected records
      may, plan ten groups instead, \texttt{edge.gof(fit, G = 10)}: they tolerate about ten exaggerated
      or five reversed records in $1000$ and ten reversed records in $5000$, at a cost of $0.017$ in mean
      power and about half the power against misfit confined to the extreme risks. Decide this before
      seeing the results. Whichever setting decides, if only the default partition rejects, check the
      records in the extreme groups. If the suspected errors are reversals only, up to about ten in
      $1000$, the symmetric basis at the default partition holds further; it fails against many
      exaggerated records. If protection matters more than power, use the Hosmer--Lemeshow statistic at
      ten groups. Beyond a few tens of corrupted records, clean the data; influence diagnostics do not
      find them (Section~\ref{sec:whatprotects}).
\item When the departure feared is a symmetric change in both tails, use the symmetric basis; on clean
      data with high discrimination (an area under the curve near $0.9$) or a strongly skewed
      covariate, in its score form, \texttt{edge.gof(fit, basis = "sym", weights = "score")}.
\item When a frozen model, logistic or not, is validated on new patients, use the external mode at ten
      groups, in which Proposition~\ref{prop:bounded} holds exactly because nothing is refitted:
      \texttt{run.all.external(y, p)}, which reports it beside Cox's recalibration test and the
      calibration intercept and slope. In the design studied its average power was that of Cox's test
      and its false-alarm rate stayed below $0.10$ where that test's did not; Stukel's and the cubic
      tests on the frozen linear predictor broke in the cohort experiment of Section~\ref{sec:realcorrupt}.
\end{enumerate}

\subsection{When another test is better}\label{sec:whennot}

Table~\ref{tab:whennot} lists the settings in which the study found a better choice. In one of them
the remedy is a different form of EDGE; in the others it is a different test. For each,
we compared the best of EDGE's forms with the rival named.

\begin{table}[htbp]\centering
\caption{Settings in which another test, or another form of EDGE, is the better choice.
Power comparisons are on clean data; the false-alarm rates quoted in the first two rows, and the last
two rows, are under corrupted records.}
\label{tab:whennot}
\small
\begin{tabular}{>{\raggedright\arraybackslash}p{0.26\textwidth}>{\raggedright\arraybackslash}p{0.21\textwidth}>{\raggedright\arraybackslash}p{0.44\textwidth}}\toprule
Setting & Better choice & Evidence \\ \midrule
Scarce events ($12\%$), clean data & GiViTI belt & The belt leads EDGE by up to $0.152$;
but it reads $0.089$ false alarms at one corrupted record in $1000$ and $0.397$ at ten. \\
A single smooth bend, clean data & Cubic calibration test or GiViTI belt & The cubic test leads in $15$
of $21$ asymmetric-link scenarios (mean $0.741$, belt $0.744$, EDGE $0.669$; largest gap $0.208$); but it
reads $0.127$
and $0.482$ false alarms at one and ten corrupted records in $1000$, where EDGE reads $0.056$ and
$0.097$. \\
A curve that crosses the diagonal twice & BAGofT or le Cessie's test & BAGofT rejects $155$ of $200$
replicates, EDGE $18$. \\
Rough or oscillating misfit & BAGofT & $1.000$ and $0.805$ on the fastest oscillation and the sawtooth,
against $0.438$ and $0.093$ for the best form of EDGE, the Stukel basis. \\
An omitted interaction & Le Cessie's test & Leads in $11$ of $15$ binary and $8$ of $12$ continuous
interaction scenarios. \\
Symmetric tails, or a strongly skewed covariate & Symmetric basis, score form & Stukel's one-parameter
score leads the default basis by $0.062$ on symmetric tails and the unit symmetric form by $0.252$ under
a skewed covariate; the score form ties it on both ($-0.002$ on average over $32$ scenarios, $-0.007$
to $+0.012$ over the four skewed ones, where its level is $0.049$ to $0.057$). \\
Many reversed records, protection first & Hosmer--Lemeshow at ten groups & $0.099$ against $0.182$ for
EDGE at ten groups with ten reversed records in $1000$, and $0.126$ against $0.249$ with
twenty-five in $5000$; its power is lower. \\
Tens of corrupted records & Clean the data & At fifty in $5000$ the default reads $0.740$ and ten groups
$0.107$; at a hundred no setting holds, and a robust fit does not help. \\
\bottomrule
\end{tabular}
\end{table}

\subsection{Limitations and further work}\label{sec:limits}

The study is a simulation over six departure families, with models of one to three covariates, five
in the external-validation study and twelve in one design, and one real cohort fitted on its routinely
recorded predictors. The corrupted-record results use one family of covariate error, a value multiplied by four or by
minus four, with three other mechanisms reported beside it; errors correlated with the outcome were not
examined. Robust tests were compared only in the form of Stukel's test and at $n=1000$, and the weighted grouped
tests of \citet{hosmer2002goodness} only with their general-purpose weight. Proposition~\ref{prop:bounded} holds at
fixed coefficients; when the model is refitted, the influence of a corrupted record through the fit is
that of the maximum-likelihood estimator, which is measured here and not bounded. And the number of
corrupted records the test tolerates is read from the measured surface: we had expected the protection
at the default partition to hold over the whole range $0.1\%$ to $0.5\%$, and at the upper end it does
not.

In external mode the test already applies to any model that reports probabilities; in the
development data the estimation adjustment $\Omega$ is derived here for logistic regression only, and
its form for other models is left for further work. A quantity that predicts the whole partition surface of
Figure~\ref{fig:partition} is still missing. The difference between wrong predictions and wrong
outcomes suggests reading a grouped and an ungrouped test together, since their disagreement carries
information about which kind of error is present. A frozen model can be monitored as patients arrive (Section~\ref{sec:stream}); a sequential reference
that uses the dependence between looks, rather than spending the level over them, is left for further
work.

%% file: sections/09_appendix.tex
\appendix

\section{Assumptions and proofs}\label{app:proofs}

This appendix proves Theorems~\ref{thm:null} and~\ref{thm:growingG}. The argument specialises two
standard results, the central limit theorem for the grouped goodness-of-fit score vector
\citep{Moore1975} and the weighted chi-squared limit of Neyman smooth components
\citep{raynerthasbest2009}, to the exact grouped logistic covariance
$\Omega=I_G-U(X^\top WX)^{-1}U^\top$ and the fixed shape basis $Z$.

\paragraph{Assumptions} Theorem~\ref{thm:null} uses (A1)--(A4) and Theorem~\ref{thm:growingG} uses
(A2), (A3) and (A4$'$). \textbf{(A1)} The number of groups $G$ is fixed and equal-frequency grouping
gives $n_g/n\to1/G$ for every $g$. \textbf{(A2)} The covariates are independent and identically
distributed with $\mathbb{E}\|x\|^2<\infty$ and $n^{-1}X^\top WX\to J_0$ positive definite, so that
the maximum-likelihood expansion $\hat\beta-\beta=(X^\top WX)^{-1}X^\top(y-\pi)+o_p(n^{-1/2})$
holds. \textbf{(A3)} The true probabilities satisfy $\varepsilon\le\pi_i\le1-\varepsilon$ for a fixed
$\varepsilon>0$, so that $V_g/n_g$ is bounded away from zero. \textbf{(A4)} The distribution of the
linear predictor is continuous with positive density at its $G-1$ population quantiles, so that the
sample group boundaries converge to fixed population boundaries and the grouping map may be treated
as fixed: each boundary is an empirical quantile fluctuating by $O_p(n^{-1/2})$, so $O_p(\sqrt n)$
observations change group; they lie in a shrinking neighbourhood of the boundary where the summands
$y_i-\hat\pi_i$ are centred to first order, so their sum is $O_p(n^{1/4})$ and, after division by
$\sqrt{V_g}=O(\sqrt n)$, contributes $o_p(1)$ to each coordinate of $r$. This is the random-cell
argument of \citet{Moore1975} for the equal-frequency partition on the fitted risk. It is used only at
fixed $G$: once $G$ grows, $\sqrt{V_g}=O(\sqrt{n/G})$ and the count no longer closes, which is why the
proof of Theorem~\ref{thm:growingG} replaces the group indicator by a smooth weight of the rank and
(A4) by \textbf{(A4$'$)}: for every $b$ in a neighbourhood of $\beta$ the distribution of $x^\top b$ is
absolutely continuous with density bounded above uniformly in $b$, and at $b=\beta$ the density of
$\eta$ is continuous and bounded away from zero on its support, which (A3) makes compact. Then
$F_\eta$ is a $C^1$ bijection of that support onto $[0,1]$, $q=F_\eta(\eta)$ is uniform, and the
quantile functions $Q_\eta$ and $Q_\pi=\mathrm{expit}\circ Q_\eta$ are $C^1$ with bounded
derivative. At fixed $G$, (A4$'$) implies (A4). Assumption (A3) restricts the asymptotic regime and
not the method; the size study of Section~\ref{sec:size} covers designs outside it.

\paragraph{Notation} Write $\varepsilon_i=y_i-\pi_i$, $v_i=\pi_i(1-\pi_i)$, $s=X^\top(y-\pi)$ for
the score, $A$ for the $G\times n$ group-membership matrix and $D=\mathrm{diag}(V_1,\dots,V_G)$, so
that the grouped residual before fitting is $\tilde r=D^{-1/2}A(y-\pi)$. For
Theorem~\ref{thm:growingG}, $q_i=F_\eta(\eta_i)$ is the population quantile position of observation
$i$, $R_i$ its rank among the $n$ fitted linear predictors, $g(i)$ its group, $q_g=(g-\tfrac12)/G$
the position of group $g$, $v(q)$ the value of $\pi(1-\pi)$ at position $q$ and
$\kappa(q)=\sqrt{v(q)}\,\mathbb{E}[x\mid q]$; $C$ is a constant not depending on $n$ or $G$.

\subsection{Proof of Theorem~\ref{thm:null}}

\paragraph{Step 1: linearisation} A one-term Taylor expansion of $\hat\pi$ about $\pi$, with the
expansion of (A2), gives $r=\tilde r-U(X^\top WX)^{-1}s+o_p(1)$ with $U=D^{-1/2}AWX$, whose rows
$U_{g\cdot}=V_g^{-1/2}\sum_{i\in g}\hat\pi_i(1-\hat\pi_i)x_i^\top$ are those of
\eqref{eq:omega}; by (A4) the grouping map may be treated as fixed.

\paragraph{Step 2: the central limit theorem} The pair $(\tilde r,(X^\top WX)^{-1/2}s)$ is a
triangular array of independent contributions, one per observation. Under (A1) and (A3) each
standardised summand is uniformly $O(1/\sqrt{\min_gV_g})=o(1)$, so the Lindeberg condition holds and
the pair is jointly asymptotically normal with $\mathrm{Cov}(\tilde r,s)=U$, each observation
belonging to exactly one group. The linear map $(a,b)\mapsto a-U(X^\top WX)^{-1/2}b$ then gives
$r\rightsquigarrow N(0,\Omega)$, since
$\mathrm{Var}(r)\to I_G-2U(X^\top WX)^{-1}U^\top+U(X^\top WX)^{-1}U^\top=\Omega$. The intercept score
equation forces $\sum_g\sqrt{V_g}\,r_g=\sum_i(y_i-\hat\pi_i)=0$ exactly, so $\Omega$ is singular in
the direction $(\sqrt{V_1},\dots,\sqrt{V_G})^\top$: a pure level shift is absorbed by the intercept.

\paragraph{Step 3: the quadratic form} By the continuous-mapping theorem
$S=r^\top P_Zr\rightsquigarrow\zeta^\top\Omega^{1/2}P_Z\Omega^{1/2}\zeta$ with $\zeta\sim N(0,I_G)$.
Let $M_\Omega=\Omega^{1/2}P_Z\Omega^{1/2}=\sum_j\lambda_je_je_j^\top$ with orthonormal $e_j$. Since
$P_Z$ has rank $d$, $M_\Omega$ has at most $d$ non-zero eigenvalues, and they are the non-zero
eigenvalues of $(Z^\top Z)^{-1}Z^\top\Omega Z$, because $\Gamma_1\Gamma_2$ and $\Gamma_2\Gamma_1$ share their
non-zero eigenvalues, with $\Gamma_1=\Omega^{1/2}Z$ and $\Gamma_2=(Z^\top Z)^{-1}Z^\top\Omega^{1/2}$. The weights lie in
$[0,1]$ because $0\preceq M_\Omega\preceq\Omega\preceq I_G$: the upper bound holds because the term
subtracted in $\Omega$ is positive semi-definite, and $\Omega\succeq0$ because $U^\top U\preceq
X^\top WX$, the between-group weighted second moment of the design never exceeding the total, by the
Cauchy--Schwarz inequality within each group. With $u_j=e_j^\top\zeta\sim N(0,1)$ independent across
$j$, $S\rightsquigarrow\sum_j\lambda_ju_j^2=\sum_j\lambda_j\chi^2_{1,j}$. The weights reported are
computed from $\widehat\Omega$; under (A1)--(A3) it differs from $\Omega$ by $O_p(n^{-1/2})$ and
Slutsky's theorem gives the same limit. \qed

\subsection{Proof of Theorem~\ref{thm:growingG}}

Let $\varphi=(\varphi_1,\dots,\varphi_d)^\top$ be the shape functions of the theorem. We show
\begin{equation}\label{eq:growingG}
S\rightsquigarrow\sum_{j=1}^d\lambda_j\chi^2_{1,j},\qquad \{\lambda_j\}=\text{eigenvalues of }
M^{-1}\Sigma,\quad \Sigma=M-KJ_0^{-1}K^\top,
\end{equation}
with $M=\int_0^1\varphi\varphi^\top dq$ and $K=\int_0^1\varphi\,\kappa^\top dq$, which do not depend
on $G_n$.

The limit is proved in six steps in Supporting Information S11: the bases are reduced to the
shape functions $\varphi$; the statistic is written as a smooth-weighted sum of the residuals;
the second-order remainder is bounded; a central limit theorem is applied with population
weights; the weights of the statistic are shown to be uniformly close to those, which is where
the rate condition $G_n/\sqrt{\log n}\to\infty$ enters; and the fitting correction gives $\Sigma$.
\qed

\subsection{Proof of Proposition~\ref{prop:bounded}}

Write $w(\eta)=\pi(\eta)\{1-\pi(\eta)\}$ with $\pi=\operatorname{expit}$. Throughout, the coefficients
are held at $\hat\beta$, as the proposition states.

\emph{The ungrouped score.} The contaminated record contributes $z(\eta^{*})(y_i-\hat\pi^{*})$ to
$U$ and $z(\eta^{*})^{2}w(\eta^{*})$ to the information. With $z(\eta)=\eta|\eta|$ and $y_i=0$ the
first is $-\eta^{*2}\hat\pi^{*}\to-\eta^{*2}$, while the second tends to zero because $w(\eta^{*})$
decays exponentially in $|\eta^{*}|$. Hence $T\ge(\eta^{*2}-c)^{2}/I_{0}\to\infty$, with $I_{0}$ the
information of the uncontaminated records and $c$ their score contribution. The same holds for the
two half-columns of the joint form and, with a higher power of $\eta$, for a polynomial of degree
four in the logit.

\emph{The grouped statistic, with the record in place.} The record enters its group only through
$E_g=\sum_{i\in g}\hat\pi_i$ and $V_g=\sum_{i\in g}w_i$, and enters the basis only through the
group's mean risk. Replacing $\hat\pi_i$ by $\hat\pi^{*}$ changes $E_g$ by $|\hat\pi^{*}-\hat\pi_i|<1$
and $V_g$ by at most $\sup_p p(1-p)=1/4$, and moves the group's mean risk by at most $1/m$. Write
$D_g=O_g-E_g$, so that $r_g=D_gV_g^{-1/2}$, and let $D_g'=D_g-\Delta E_g$ and $V_g'=V_g+\Delta V_g$ be
the values after the change; $O_g$ does not change, because the outcome does not. Then
\[
  r_g'-r_g=-\frac{\Delta E_g}{\sqrt{V_g'}}+D_g\Bigl(\frac1{\sqrt{V_g'}}-\frac1{\sqrt{V_g}}\Bigr).
\]
For $V_g>1/4$ both $V_g$ and $V_g'$ are at least $V_g-1/4$, and by the mean value theorem
$|V_g'^{-1/2}-V_g^{-1/2}|\le\tfrac12|\Delta V_g|(V_g-\tfrac14)^{-3/2}\le\tfrac18(V_g-\tfrac14)^{-3/2}$.
With $|D_g|=|r_g|\sqrt{V_g}$ this gives the bound of the proposition,
\[
  |\Delta r_g|\;\le\;\Bigl(V_g-\tfrac14\Bigr)^{-1/2}
  \Bigl\{1+\frac{|r_g|\sqrt{V_g}}{8\,(V_g-\frac14)}\Bigr\}.
\]
The basis columns are scaled to unit length and are Lipschitz in the group mean, with a constant that
for the logit-based columns depends on the extreme groups' mean risk but not on $\eta^{*}$, so the change they
contribute to $S=r^\top P_Z r$ is bounded by a constant multiple of $1/m$. No term involves
$\eta^{*}$.

\emph{The grouped statistic, when the record changes group.} Equal-frequency grouping moves the
corrupted record into an extreme group and displaces one record across each boundary between. Each
touched group loses one record and gains one, so $O_g$ changes by at most one, $E_g$ by less than one
and $V_g$ by at most $1/4$. The argument above, with $|\Delta D_g|\le2$ in place of $|\Delta E_g|<1$,
gives $|\Delta r_g|\le(V_g-\frac14)^{-1/2}\{2+|r_g|\sqrt{V_g}/(8(V_g-\frac14))\}$ for every touched
group, again with no dependence on $\eta^{*}$.

At fixed coefficients the influence of one record's covariate on the grouped statistic is therefore
bounded by a sum, over the groups the record touches, of quantities that do not depend on $\eta^{*}$,
while its influence on the ungrouped directed score is
unbounded. \qed